UNIVERSITY OF CALIFORNIA,
IRVINE

Symmetries, Dark Matter and Minicharged Particles

DISSERTATION

submitted in partial satisfaction of the requirements
for the degree of

DOCTOR OF PHILOSOPHY

in Physics

by

Jennifer Rittenhouse West

Dissertation Committee:
Professor Tim Tait, Chair
Professor Herbert Hamber
Professor Yuri Shirman

2019

# DEDICATION

To my wonderful nieces & nephews, Vivian Violet, Sylvie Blue, Sage William, Micah James &
Michelle Francesca
with all my love and so much freedom for your beautiful souls
To my siblings, Marlys Mitchell West, Jonathan Hopkins West, Matthew Blake Evans
Tied together by the thread in Marlys's poem, I love you so much and I see your light
To my mother, Carolyn Blake Evans, formerly Margaret Carolyn Blake, also Dickie Blake,
Chickie-Dickie, Chickie D, Mimikins, Mumsie
For never giving up, for seeing me through to the absolute end, for trying to catch your siblings in
your dreams. I love you forever.
To my father, Hugh Hopkins West, Papa-san, with all my love.
To my step-pop, Robert Joseph Evans, my step-mum, Ann Wilkinson West, for loving me and
loving Dickie and Papa-san.
In memory of Jane.
My kingdom for you to still be here. My kingdom for you to have stayed so lionhearted and sane.
I carry on. Piggy is with me.
In memory of Giovanni, with love and sorrow.
Io sono qui.
Finally, to Physics, the study of Nature, which is so deeply in my bones that it cannot be removed.
I am so grateful to be able to try to understand.
...
*"You see I am absorbed in painting with all my strength; I am absorbed in color - until now I have
restrained myself, and I am not sorry for it. If I had not drawn so much, I should not be able to
catch the feeling of and get hold of a figure that looks like an unfinished terracotta. But now I feel
myself on the open sea - the painting must be continued with all the strength I can give it."*
...
*- Vincent van Gogh to his brother Theo, 3 September 1882*

# TABLE OF CONTENTS

# LIST OF FIGURES

# LIST OF TABLES

# ACKNOWLEDGMENTS

This dissertation research was partially supported by the Graduate Assistance in Areas of National Need (GAANN) (Award No. P200A150121) Federal Fellowship. I would like to thank Prof. Asantha Cooray for awarding me this fellowship.

I am grateful to the University of California, Irvine DECADE program for $4$ years of summer funding and personal support.

The final year of my Ph.D. work was carried out at SLAC National Accelerator Laboratory and I am infinitely grateful to Michael Peskin, Stan Brodsky and Tom Rizzo for my space here. I thank Alexander Friedland for his brilliant and inspiring neutrino physics class at Stanford University which sustained me for months while writing this thesis, besides teaching me everything I know about neutrinos. I am honored to be his colleague and consider him my friend. We are both dinosaurs in a sense, with very old school notions of what physics is and the importance of curiosity.

Professors Stan Brodsky and James Daniel Bjorken at SLAC have become my collaborators over this past year and it turns out that working with legends - who are not only brilliant but very direct, very kind - is good for the soul as well as for physics. I thank them both from my heart. I consider them both my friends, as well as Stan's wife Judy Brodsky. The Brodsky dinners and hospitality are wonderful. They make everyone feel welcome and I needed very much to feel welcome here.

I thank Herbert Hamber for his kindness and for teaching me the beautiful theory of general relativity. I thank both Herbert and his wife Franca for their darling parties up on Turtle Hill, for being so kind and welcoming to me. Herbert was my first advisor before I switched subfields from quantum gravity to particle theory and he graciously stayed on my thesis committee afterwards as well as serving as my unofficial committee chair. Thank you, Herbert.

The first research advisors I ever had were at the University of California, Santa Barbara, Professors France Anne-Dominic Córdova and James Hartle. I began graduate studies at UCSB doing research with the Córdova Optical Monitor group, a part of the European Space Agency's XMM-Newton telescope group. My x-ray astrophysics work with France was supplemented by observing runs in the optical band at Lick Observatory in northern Calfornia and I am grateful to her for sending me to Mount Hamilton for many long nights, often alone. She took me very seriously. Anytime I had a science idea - e.g. to jump on the underutilized Nickel 1-meter telescope at Mount Hamilton in order to get optical followup a possible black hole binary system I'd read about in an astronomy telegram - France immediately supported my wish to go with funding and optimism. I didn't realize until much later how important this is, to be taken so seriously and to be fully funded, and also how rare. I thank France from my heart.

Jim Hartle was my undergraduate advisor in generalized quantum mechanics research and working with him was an honor. He was so supportive and kind to me. I will always be grateful to him for this early experience. I worked with Doug Eardley during those same years, on a time delay in gravitational lensing calculation. This was another wonderful experience as Doug's combination of incisiveness and patience made me feel valued and capable.

I must include my gratitude to Professor Virginia Louise Trimble who has been a generous colleague and confidante during my time at UCI. She and I have painfully sharp differences of opinion regarding women in physics and astronomy. I believe that harassment of women and minoritized people in any form is wrong and punishable and that abusers must be held to account and Virginia, with her Ph.D. from Caltech in the late 1960s, does not. I discovered this the year I moved up to SLAC, soon after I co-organized a 2018 symposium in her honor with the most wonderful co-organizers a person could ask for - Martin Rees, Lynn Cominsky & David Helfand. I deeply thank these three. We really went through it and all of us showed our true colors, complementary colors indeed.

I thank Virginia for trusting me to substitute teach her classes, for introducing me to wonderful people, great scientists and great non-scientists, and for being so supportive to me while I was in southern California. It does not surprise me that my ex-advisor France Córdova wanted her colleague V.L.T. at her inauguration as the director of the National Science Foundation. Virginia and France went through the fire together in my opinion, though eleven years apart for their Ph.D.s at Caltech. Their mutual admiration and support is a comfort to me.

This dissertation and defense would have gone much differently if it weren't for one of the most wonderful people I've ever known, Professor Isabella Velicogna. Her life is extremely busy with family, faculty and associate dean duties but she made time for me when she heard of my situation at UCI. We were hundreds of miles apart but she was on the phone with me, texting and calling, whenever I needed her. The night before my thesis defense I was out walking the hills and paths of SLAC crying on the phone with Isabella, past midnight and truly frightened. I flew into town the next morning to defend and came directly to her building. Her husband Eric Rignot, professor and chair of the Earth Sciences department, is cut from the same deeply kind cloth which I felt in an instant. I am so grateful to both of them. I am happy to acknowledge Isabella Velicogna's contributions to my work and my life and I love her for what she did for me. She is not only brilliant but forthright and giant hearted, a true friend to me and ally to women in science.

I have dedicated my thesis to my family and to physics. I will only mention again my sister, Marlys West, a sister of sisters, brave as a thousand lions and bold and brilliant too. I acknowledge old and deep debts to her with infinite gratitude.

# CURRICULUM VITAE

## Jennifer Rittenhouse West

**EDUCATION**

| | |
|---|---:|
| **Doctor of Philosophy in Physics** | **2019** |
| University of California Irvine | *Irvine, CA* |
| **Masters of Science in Physics** | **2012** |
| University of California Davis | *Davis, CA* |
| **Bachelor of Science in Physics** | **1998** |
| University of California Santa Barbara | *Santa Barbara, CA* |

**RESEARCH EXPERIENCE**

| | |
|---|---:|
| **Visiting Ph.D. Candidate** | **2018–Present** |
| SLAC National Accelerator Laboratory | *Menlo Park, California* |
| **Graduate Research Assistant** | **2015–2018** |
| University of California, Irvine | *Irvine, California* |
| **X-ray Astronomy Research and Chandra Ground Support** | **2002–2008** |
| Harvard-Smithsonian Center for Astrophysics | *Cambridge, Massachusetts* |

**TEACHING EXPERIENCE**

| | |
|---|---:|
| **Teaching Assistant** | **2013–2015** |
| University of California Irvine | *Irvine, CA* |
| **Teaching Assistant** | **2009–2012** |
| University of California Davis | *Davis, CA* |

# REFEREED JOURNAL PUBLICATIONS

**Millicharged Scalar Fields, Massive Photons and the Breaking of $SU(3)_C \times U(1)_{\text{EM}}$**  April 2019
Physical Review D 99, 073009

**Asymmetric Dark Matter and Baryogenesis from $SU(2)_\ell$**  August 2017
Physical Review D 96, 035001

**Chandra Observations of the Gamma-ray Binary LSI+61303: Extended X-ray Structure?**  July 2007
The Astrophysical Journal Letters, Volume 664, Number 1

**First XMM-Newton observations of a cataclysmic variable I: Timing studies of OY Car**  January 2001
Astronomy & Astrophysics 365, L288-L293

# CONFERENCE PROCEEDINGS

**Statistical Challenges in the Search for Dark Matter**  2018
Banff International Research Station Dark Matter Astroparticle Workshop

# TALKS & SEMINARS

**"Millicharged Fields in Particle Physics and Cosmology"**  February 2019
Particle Theory seminar, SLAC National Accelerator Laboratory

**"Multi-purpose Millicharged Fields: Dark Matter, Colliders & Cosmology"**  April 2019
Particle Theory seminar, College of William & Mary

# ABSTRACT OF THE DISSERTATION

Symmetries, Dark Matter and Minicharged Particles

By

Jennifer Rittenhouse West

Doctor of Philosophy in Physics

University of California, Irvine, 2019

Professor Tim Tait, Chair

This theoretical particle physics thesis is an investigation into old and new symmetries of Nature. Known symmetries and conservation laws serve as a guide for dark and visible sector model building. New symmetries of Nature are proposed, broken and subsequently reinstated at high temperatures in order to discover well-motivated particle physics models for cosmological observations implying the existence of a dark sector. Candidate processes for creation of a non-primordial matter/antimatter asymmetry result from out of equilibrium spontaneous breaking of these symmetries in the early Universe. Using the Standard Model of particle physics as a foundation with minimal new degrees of freedom, minicharged and millicharged particles emerge from a proposed spontaneous breaking of known symmetries. Experimental predictions and constraints for such dark matter candidates are given briefly here and outlined for future work. Constraints on neutrino-like particles found in the debris of broken local (gauge) symmetries are given, a subset of which are sterile and appear to be viable particle dark matter candidates. A failed baryonic dark matter candidate became a candidate to solve an outstanding nuclear structure problem, the EMC effect.

# Chapter 1

# Introduction

This thesis is concerned with symmetries of Nature both old and new. The study of modern theoretical particle physics and cosmology has been built upon the symmetries that the underlying equations of physics obey. Why should this be so? It is so because there is a deep connection between symmetries and conserved quantities in Nature. The brilliant German mathematician Emmy Noether (1882-1935) was the first to realize this connection and the resulting theorem bears her name [1]. Nature contains special quantities which do not change, regardless of the processes they undergo. Finding and studying quantities that Nature treats this way is one of the most powerful paths forward in theoretical physics, from the known to the unknown.

Experiments show that there are charges in particle physics that are conserved: electric charge, strong force color charge, weak force charge. By Noether's theorem, each charge or current is related to a symmetry of the Lagrangian, $\mathscr{L} = \mathrm{T} - \mathrm{V}$, the difference between the kinetic and potential energy of a system. For the three known and experimentally verified particle interactions, these symmetries are quantum mechanical and are described by renormalizable quantum field theories. They are local in the sense that the symmetry transformations are spacetime dependent; an observer on Earth may transform her equations by a different amount than an observer in the far future in the Whirlpool galaxy and still their invariant quantities will be identical.

One of the four known interactions does not follow this script. The gravitational force obeys a classical local symmetry. Gravity obeys the conservation of energy[1] and momentum by insisting upon equations invariant under local Poincaré transformations,

$$\Delta x^{\overline{\alpha}} = \sum_{\beta=0}^{3} \Lambda_{\beta}^{\overline{\alpha}} \Delta x^{\beta}, \text{ for arbitrary } \overline{\alpha} \tag{1.1}$$

where $\Delta \vec{x} \rightarrow (\Delta t, \Delta x, \Delta y, \Delta z)$ is the displacement vector ($\Delta \vec{x} \rightarrow \{\Delta x^{\alpha}\}$ in tensor notation), and

$$\Lambda_{0}^{\overline{\alpha}} = 1/\sqrt{(1-v^2)}, \quad \Lambda_{1}^{\overline{\alpha}} = -v/\sqrt{(1-v^2)}, \quad \Lambda_{2}^{\overline{\alpha}} = \Lambda_{3}^{\overline{\alpha}} = 0 \tag{1.2}$$

are the Lorentz transformations from the unbarred to the barred coordinate systems. A boost such as this along a direction labeled $x$ in Minkowski space leaves all physical laws invariant [3]. Thus observers traveling in different reference frames, as, for example, a physicist on Earth at $x^{\beta}$ and a physicist living in the Eridanus II stellar cluster at $x^{\overline{\alpha}}$, find that the same physical laws hold, the same quantities are conserved. This holds for translations, rotations and velocity boosts between inertial frames. However, the gravitational force as described by Einstein's theory of general relativity [4] is the only one of the four known forces that has not been discovered to be a viable local quantum symmetry of Nature. It is nonrenormalizable and not all physical quantities are calculably finite. The renormalization of gravity is a vast and rich field of theoretical physics. It may, if discovered, turn out to be related to the symmetry breaking and restoration techniques used throughout this thesis and be described by spacetime dependent local transformations just like particle interactions. As of today this has not been shown to hold. Therefore, the gravitational interaction will not be discussed further except as a classical gravitational field [4] whose equations of motion imply the existence and behavior of dark matter and dark energy, two of the most mysterious phenomena in physics today.

---

[1]On cosmological scales in Einstein's general relativity the energy density changes as space itself is expanding. [2], however the divergence of energy-momentum tensor is always zero and energy-momentum is conserved.

The foundation of the thesis is charges and currents known to be conserved in the Standard Model of particle physics [5]. With one notable exception - nonperturbative interactions induced by solutions to the classical equations of motion for finite Euclidean action $S = \int \mathscr{L} \mathrm{d}^4 \mathrm{x}$ [6], known as instantons - modeling particle physics phenomena with conserved currents may be conducted in the following manner: Determine the interactions of interest. Let us say, for example, we are concerned with the force of electromagnetism, as indeed Chapter 2 is intimately concerned with. Notice and verify that all experiments demonstrate a conserved quantity, here the electric 4-current $j_\mu = (\rho, \mathbf{j})$ where $\rho$ is the charge density $\frac{q}{V}$ and $\mathbf{j}$ is the 3-current density, $\frac{\mathbf{J}}{V}$. If the interaction is known and experimentally verified, simply list the degrees of freedom, i.e. the particles and fields participating in the interaction. All electrically charged particles feel this force and one gauge boson, the photon, mediates it. An interaction such as quantum electrodynamics (QED) with one massless gauge boson is described by the local symmetry $e^{iq\theta(x)}$ acting upon all of the participating fields, the photon field $A_\mu$, and the charged fermionic fields $\psi(x_\mu)$. Build a Lagrangian with every possible combination of such fields such that every term in $\mathscr{L}$ is invariant under the symmetry transformation. This uniquely specifies the Lagrangian invariant under the symmetry group of unitary transformations $e^{iq\theta(x)}$.

The Lagrangian thus obtained may be used to discover and predict new physical phenomena. All of Maxwell's equations of electromagnetism are derivable from the $U(1)_{\mathrm{EM}}$ Lagrangian by varying and extremizing the resulting action $\delta S_{\mathrm{EM}} = \delta \left\{ \int dt \, \mathscr{L}_{\mathrm{EM}} \right\} = 0$ and finding the equations of motion [7].

Any physical phenomena not arising from the manipulations of $\mathscr{L}_{\mathrm{EM}}$ may be considered forbidden (again with the notable exception mentioned previously [6] and made use of in Chapters 3 and 4). Physical phenomena allowed by $\mathscr{L}_{\mathrm{EM}}$ may be considered mandatory. Many experiments have been and currently are built upon these two considerations. This is the power of Noether's theorem. The published work of Chapter 2 reflects ancillary research of the known application of Noether's theorem to electromagnetism.

It can be used as an equally powerful tool in the opposite direction. Chapter 3 of this thesis is concerned with exactly this behavior. In this case one does not have definitive experimental results with the known gauge bosons (force mediators) and charged particles needed to construct a symmetry that can predict new physics results. Even the conserved charges, if they exist, are unknown. What is known is indirect evidence for mysterious phenomena in Nature. The current epoch of physics overflows with such phenomena, some of which will be described in the following section and two of which have already been mentioned. In this section, the procedure for uncovering physical laws that predict experimental results is described just as was done for the case of experiment leading the way to symmetries. If there is evidence for new phenomena, for example the gravitational evidence for the existence of dark matter, this procedure may be of great use. It must be noted that we do not know if dark matter is a particle obeying new or old local quantum symmetries. It may be a purely gravitational phenomena. It may be related to geometry in a deep and complex way that has not been revealed. ² In this thesis particle dark matter is assumed, as it is by most physicists working in the field of phenomenological particle physics today, but this may turn out to be an incorrect assumption.

The gravitational effects indicating the existence of dark matter have been extremely well studied [8] beginning with Fritz Zwicky's work on the Coma Cluster of galaxies in 1933 [9]. Gravitational rotation curves, Big Bang nucleosynthesis, the fluctuations in the Cosmic Microwave Background radiation (CMB), the colliding clusters of galaxies known as the Bullet Cluster, all of these and many more give strong gravitational evidence for another form of matter which has been so far completely invisible to us via any other type of interaction. The search for the non-gravitational evidence of dark matter is a vigorous research area in both theoretical and experimental particle physics [10]. As of today, no $5\sigma$ reproducible results have illuminated the identity of dark matter.

---

²As a notable example, James D. Bjorken is working on an extra dimensional theory with a relaxed spacetime isotropy condition at SLAC National Accelerator Laboratory. I am honored to be working with him on a separate project involving dark energy dominated cosmic voids. His goal is to have physical phenomena with strong observational evidence - here, the uniformity of the temperature of the Cosmic Background Radiation (CBR) to within one part in $10^5$ - arise from the geometry of the extra dimensions behaving only under the force of gravity.

In this case, model building would not begin with known conserved charges and the masses and numbers of gauge bosons which point to a symmetry for the equations. The methodology for dark matter is often to guess a possible symmetry for the equations to obey and then work out the fields necessary for the symmetry to be local (i.e. cancel gauge anomalies given in Feynman's language by the triangle diagrams [11]), finally deriving the conserved quantities from the Lagrangian built upon the hypothesized symmetry. These new interactions and particles are then matched to the cosmological behavior of dark matter (or e.g. dark energy, or inflation) and if the match is a good one, the Lagrangian is inspected for any new phenomena that may be predicted and experimentally verified. Ideally the guesses are well motivated when lacking non-gravitational experimental constraints. "Well motivated" here means making use of known physical behavior. Nature is known to obey local and global symmetries, with local symmetries giving rise to interactions or forces. This feels like a good starting place to approach the unknown. As with all good starting positions, it may in later years become necessary to leave it behind. One primary reason for this comment is the mismatch between constructing theories in the flavor basis - this is the symmetry basis between generations of particles - and measuring observables in the mass basis. The fact that the flavor basis is not diagonal and therefore not equivalent to the physical basis is highly suggestive of new physics. One possible pathway for new physics would be to postulate the non-fundamental nature of local symmetries. I confess I would not like this path but I would follow it if experiment showed it to hold true in any regime.

The dark matter and baryogenesis model of Chapter 3 was an attempt to utilize this procedure of guessing new symmetries and building Lagrangians upon the hypotheses. In Chapter 4, which describes both work in progress and future work, a perhaps better motivated dark matter and baryogenesis model is offered. The difference between the models is in their quantum behavior. Chapter 3 is based upon an anomalous global symmetry, the related work in progress outlined in Chapter 4 is built upon a symmetry that holds both globally and locally. That symmetry is baryon number minus lepton number, $B - L$.

An in depth analysis of the breaking of old and new symmetries follows.

# Chapter 2

# Millicharged Fields and the Breaking of Symmetries

Based on arXiv:1711.04534v4 [12]

## 2.1 Overview

This work began by considering the symmetry structure of the Standard Model of particle physics in the past and realizing that mirror processes may happen in the future. It is perhaps natural to consider the space and time of here and today as somehow static or pinned down, when in fact we may exist in a dynamical situation not only in regards to the expansion of the Universe but also with respect to the local symmetry structure. The implications of discarding this natural consideration are investigated here.

The assumption that the current epoch of the Universe is not special, i.e. is not the final state of a long history of processes in particle physics, allows the cosmological fate of $SU(3)_C \times U(1)_{\text{EM}}$ to be investigated. Spontaneous symmetry breaking of $U(1)_{\text{EM}}$ at the temperature of the Universe today is carried out. The charged scalar field $\phi_{\text{EM}}$ which breaks the symmetry is found to be ruled out for the charge of the electron, $q = e$. Scalar fields with millicharges are viable and limits on their masses and charges are found to be $q \lesssim 10^{-3}e$ and $m_{\phi_{\text{EM}}} \lesssim 10^{-5}$eV. Furthermore, it is possible that $U(1)_{\text{EM}}$ has already been broken at temperatures higher than $T = 2.7K$ given the nonzero limits on the mass of the photon. A photon mass of $m_\gamma = 10^{-18}$eV, the current upper limit, is found to require a spontaneous symmetry breaking scalar mass of $m_{\phi_{\text{EM}}} \sim 10^{-13}$eV with charge $q = 10^{-6}e$, well within the allowed parameter space of the model. Finally, the cosmological fate of the strong interaction is studied. $SU(3)_C$ is tested for complementarity in which the confinement phase of QCD + colored scalars is equivalent to a spontaneously broken $SU(3)$ gauge theory. If complementarity is not applicable, $SU(3)_C$ has multiple symmetry breaking paths with various final symmetry structures. The stability of the colored vacuum at finite temperature in this scenario is nonperturbative and a definitive statement on the fate of $SU(3)_C$ is left open. Cosmological implications for the metastability of the vacua - electromagnetic, color and electroweak - are discussed.

## 2.2   Introduction

By Noether's theorem as described in Chapter 1, symmetries of Nature are deeply connected to particles and their interactions. The work presented in this chapter builds upon this fact [12]. The fundamental symmetry structure of the Universe has changed at least once over the past $13.8 \times 10^9$ years. The early Universe combined the weak and electromagnetic interactions, a symmetry that was broken by the Higgs field and is described by the Standard Model (SM) of particle physics. The SM contains all currently known particles and interactions, with the exception of neutrino masses, as well as the spontaneous symmetry breaking (SSB) mechanism of the Higgs field.

Grand unified theories (GUTs) unify the strong and electroweak sectors of particle physics into larger symmetry groups which are typically broken down to the SM by new scalar fields at earlier times and higher temperatures [13]. They are highly motivated by physics beyond the SM, notably dark matter and quantum gravity, and if realized in Nature would extend the symmetry breaking pattern of the past.

It may be of interest to study symmetry breaking in the future. If the current age and temperature of the Universe are not special with respect to symmetries, then as the previous group structure of the Universe was broken at least once, so may the current structure be broken one or more times. In this case the fate of the Universe may be determined by studying the possible symmetry breaking paths and the effects thereof.

The particle physics framework of the $2.7~K$ Universe is the gauge group structure $SU(3)_C \times U(1)_{\text{EM}}$. The remarkable success of both QED and QCD in predicting particle properties, decays and interactions gives compelling evidence that these local symmetries hold today. At earlier times, i.e. at temperatures greater than $T \sim 100$ GeV, electroweak symmetry breaking had not yet occurred and the larger group structure of the SM, $SU(3)_C \times SU(2)_L \times U(1)_Y$, held. The hypothesis of spontaneous symmetry breaking of the SM via the scalar Higgs field was confirmed in 2012 in a stunning achievement of experimental collider physics [14, 15]. That discovery, a proof of existence in a sense, allows for the question of future SSB. For SSB to occur, new scalar fields with color and/or electromagnetic charge are needed. The shape of the the scalar potential must be such that the proposed $SU(3)_C \times U(1)_{\text{EM}}$ vacuum is metastable. This work investigates whether the current group structure is truly the final symmetry state of the Universe.

The cosmological fate of $SU(3)_C \times U(1)_{\text{EM}}$ may, under highly specific conditions, affect the fate of the Universe. If $U(1)_{\text{EM}}$ is spontaneously broken by an electromagnetically charged scalar field, the cosmic microwave background (CMB) photons gain mass. The photon mass, as shown in Section 2.4, is dependent upon the vacuum expectation value (vev) of the scalar field,

$$m_\gamma = \sqrt{2}qv, \tag{2.1}$$

where $q$ is the charge of the scalar field in units of electron charge and $v$ is the vev.

If the acceleration of the expansion of the Universe is caused by a cosmological constant, $\Lambda$, this gain in mass (regardless of the value of the vev) will have no effect on the fate of the Universe under the usual assumption of a Friedmann-Lemaître-Robertson-Walker (FLRW) Universe. In the homogenous, isotropic and flat FLRW Universe the total energy density, having previously passed through phases of radiation domination followed by matter domination, has recently entered the cosmological constant dominated phase. During this phase, nothing can overtake the effect of $\Lambda$'s energy density on the expansion rate and the expansion will be eternal and accelerated [16].

However, if the acceleration is not caused by a cosmological constant and instead dark energy evolves in the future in such a way that its energy density parameter varies as the scale factor $a^y(t)$ with $y < -3$, the CMB photons' gain in mass could be important. Radiation is the only currently known type of energy density that evolves in the necessary way, as $a^{-4}(t)$. If the dark energy were to be modeled by a scalar field (or fields) which decays in the future to radiation then the breaking of $U(1)_{\text{EM}}$ could affect the fate of the Universe given a large enough value of $m_\gamma$. The expansion rate could slow down or even reverse.

The 2018 cosmological parameters from the *Planck* satellite strongly favor a small positive cosmological constant today [8]. There is currently no reason to believe the evolution outlined above would occur, however, the cause of the acceleration of the expansion is unknown and some kind of evolution in time is plausible.

For the SSB potential considered in this work, the allowed masses for an electromagnetically charged scalar field are quite small and must be millicharged in order to be viable. More complicated scalar potentials, e.g. composed of scalar fields carrying both electromagnetic and color charge, or the use of a non-SSB mechanism (e.g. radiative symmetry breaking [17]) could affect this conclusion. Previous studies of a charged Higgs boson related to the SM Higgs at finite temperatures did not support a SSB, a fact relayed to the author after submitting an earlier version of this work [18].

It would be interesting to allow for higher mass millicharged fields as these are excellent dark matter candidates, e.g. [19]. New experiments such as the Light Dark Matter Experiment (LDMX) [20], MilliQuan [21], NA64 [22] and SHiP [23] propose to detect $\sim 1$ MeV to $\sim 10$ GeV particles with charges from $10^{-1}e$ to $10^{-4}e$. Sub-MeV millicharged particle tabletop detectors are currently in development as well. The millicharged particles discussed in this work are too light for these direct detection experiments but they may be of interest for next generation experiments. There is a possibility of pushing to higher masses while retaining millicharges, discussed in Section 2.4.5.

## 2.3 Stability of the Vacua

The implicit assumption is that the electroweak vacuum is stable and the SM Higgs vacuum expectation value of 246 GeV is the true vacuum. This may not be true. More precise measurements of the Higgs coupling to the top quark are needed to determine the stability of the electroweak vacuum, with consequences of metastability outlined in the 1980s [24]. A state-of-the-art calculation [25] suggests that we are in a metastable electroweak vacuum with a lifetime of

$$\tau = 10^{561^{+817}_{-270}} \text{ years,} \qquad (2.2)$$

with the given uncertainties due only to the top quark mass (other SM parameter measurement uncertainties contribute but the top quark mass dominates). This result may settle into absolute stability or shorter lifetime metastability by physics beyond the SM, including a theory of quantum gravity [26, 27]. The necessary precision on the top quark mass for a $3\sigma$ metastability confirmation is $\Delta m_t < 250$ MeV [28]. With the current uncertainty of the top quark mass from direct measurements, $m_t = 173.21 \pm 0.51 \pm 0.71$ GeV [29], this question will likely not be answered with additional Large Hadron Collider (LHC) Run II results. Uncertainties of less than 200 MeV may be accessible when the high luminosity (HL-LHC) upgrade is complete and the full dataset taken [30].

The stability of a QCD-QED vacuum is a separate question. The vacuum metastability investigated in this case is due to a colored scalar field or fields with a nonzero vev and/or an electromagnetically charged scalar field with a nonzero vev. The presence of a new scalar charged under $SU(2)_W \times U(1)_Y$ - as any field with $q \neq 0$ must be - affects the shape of the Higgs potential at high energies and therefore may have an effect on electroweak vacuum stability.

If it were possible to rule out the existence of a new electromagnetically charged scalar field, the electroweak vacuum stability question would remain dependent upon future precision measurements of the top quark mass as well as any new physics effects. The sub-eV mass millicharged scalar fields used in the model presented here would have very little effect on the shape of the Higgs potential but with higher mass millicharged scalar fields (discussed in Section 2.4.5) this could change.

## 2.4  $U(1)_{\text{EM}}$ spontaneous symmetry breaking

In order to break the $U(1)_{\text{EM}}$ gauge symmetry, an electromagnetically charged scalar field $\phi_{\text{EM}}$ is introduced (the subscript will be dropped for clarity). It is a color singlet with charge $q$ under $U(1)_{\text{EM}}$ and gives rise to a new scalar section of the QED Lagrangian,

$$\mathscr{L}_{QED} \supset -\frac{1}{4}F^{\mu\nu}F_{\mu\nu} + D^\mu \phi^* D_\mu \phi - V(\phi), \tag{2.3}$$

with covariant derivative $D_\nu = \partial_\nu + iqA_\nu$ and field transformations under $U(1)_{\text{EM}}$

$$\phi \to e^{i\alpha}\phi, \ A_\nu \to A_\nu - \frac{1}{e}\partial_\nu \alpha. \tag{2.4}$$

The scalar potential is given by

$$V(\phi) = -\frac{\mu^2}{2}|\phi|^2 + \frac{\lambda}{4!}(\phi^*\phi)^2 \tag{2.5}$$

where the field $\phi$ gains a vev for the choice of mass parameter $\mu^2 > 0$. Labeling the minimum of the potential $v_e$,

$$v_e^2 \equiv \langle \phi \rangle^2 = \frac{6\mu^2}{\lambda} \tag{2.6}$$

the Lagrangian is expanded about the minimum. The complex scalar field may be written as $\phi = v_e + \frac{1}{\sqrt{2}}(\phi_1 + i\phi_2)$ to yield

$$V(\phi) = -\frac{3}{2\lambda}\mu^4 + \mu^2\phi_1^2 + \mathcal{O}(\phi_i^3) \tag{2.7}$$

One of the scalar fields gains mass $m_\phi = \mu$ and the other is the massless pseudo-Goldstone boson which provides the longitudinal polarization of the now massive photon.

The photon gains its mass via the kinetic energy term of $\mathscr{L}$,

$$m_A^2 = 2q^2 v_e^2. \tag{2.8}$$

### 2.4.1 Electromagnetic scalar potential in the $2.7\,K$ Universe

In order to study the effects of new scalar fields in the current epoch, we calculate the scalar potential at the temperature of the cosmic background radiation today. Computing $V(\phi, T)$ for $T \approx 10^{-4}$ eV gives an estimate of the effects in terms of relationships between the parameters of the model.

It is important to note that the finite temperature field theory equations assume both equilibrium conditions and homogeneity of the medium. To accommodate the non-equilibrium conditions - the CMB has a thermal distribution but is not in equilibrium due to the expansion of the Universe - a time slice at $T = 2.7$ K is used. Equilibrium is assumed for this moment in time. The assumption of homogeneity in the Universe is length scale dependent. On the largest length scales both homogeneity and isotropy appear to hold. This work is concerned with such cosmological scales.

Closely following the treatment of Quirós 1999 [31] and Coleman and Weinberg 1973 [17], the finite temperature potential in terms of the constant background field $\phi_e$ is given by

$$V(\phi_e, T) = V_0(\phi_e) + V_1(\phi_e, 0) + V_{1T}(\phi_e, T), \tag{2.9}$$

where the first term is the zero temperature classical potential as in (2.5), the second term is the zero temperature Coleman-Weinberg correction to one-loop order and the final term is the finite temperature contribution, also calculated to one-loop.

Both zero temperature and finite temperature loop calculations include contributions from all relevant particles coupled to the scalar field. The gauge boson of the $U(1)_{\text{EM}}$ gauge group, the fermions charged under it, and the scalar field itself all may run in the loop. Fermions will not be relevant for the temperatures and densities considered here due to the baryon-to-photon ratio data as given by Big Bang Nucleosynthesis (BBN) [32]

$$5.8 \times 10^{-10} \leq n_{\text{b}}/n_\gamma \leq 6.6 \times 10^{-10} (95\%\text{CL}), \tag{2.10}$$

however photons and $\phi$ both contribute to the effective potential.

The 1-loop $T = 0$ contributions are given, using $\overline{MS}$ renormalization counter terms with a cut-off regularization and the assumption that the minimum and the scalar mass do not change with respect to their tree level values, that is

$$\left.\frac{d(V_1 + V_1^{c.t.})}{d\phi_e}\right|_{\phi_e=v_e} = 0 \tag{2.11}$$

and

$$\left.\frac{d^2(V_1 + V_1^{c.t.})}{d(\phi_e)^2}\right|_{\phi_e=v_e} = 0, \tag{2.12}$$

by the following equation

$$V_1(\phi_e) = \frac{1}{64\pi^2} \sum_i n_i \{m_i^4(\phi_e)(\log \frac{m_i^2(\phi_e)}{m_i^2(v_e)} - \frac{3}{2}) + 2m_i^2(v_e)m_i^2(\phi_e)\}. \qquad (2.13)$$

Here $i = \gamma, \phi$ and $n_i$ the degrees of freedom with $n_\gamma = 3$ for the newly massive photon and $n_\phi = 1$ for the scalar field.

The contributions to the thermal effective potential to 1-loop order are given by

$$V_{1T} = \sum_i \frac{n_i}{2\pi^2 \beta^4} \int_0^\infty dx\, x^2 \log\left(1 - e^{-\sqrt{x^2 + \beta^2 m_i^2(\phi_e)}}\right). \qquad (2.14)$$

The high temperature expansion cannot be used in the case of the $T = 2.7K$ Universe and an analytic solution to the temperature dependent integral does not exist. However, a numerical solution is possible under the conditions outlined in the following subsection.

### 2.4.2 Spontaneous symmetry breaking conditions

Any future spontaneous symmetry breaking will depend upon the sign of the quadratic coefficient in the effective potential, $\frac{d^2V}{d\phi_e^2}$ evaluated at $\phi_e = 0$. Two constraints must be satisfied. First, that there has been no SSB until today. This stability condition becomes

$$\left.\frac{d^2V}{d\phi_e^2}\right|_{\phi_e=0,\, T \geq 2.7K} \geq 0. \qquad (2.15)$$

Second, that a SSB may occur in the future, and let us take the furthest future possible in temperature, i.e. $T = 0$,

$$\frac{d^2V}{d\phi_e^2}\bigg|_{\phi_e=0, T=0} < 0. \tag{2.16}$$

When this second derivative is negative for temperatures $T < 2.7K$, with the additional requirement that $\lambda > 0$, SSB will occur. Setting these constraints on $V(\phi_e, T)$ allows for a numerical evaluation of the integrals for any value of $\mu^2$ and $\lambda$, as the derivatives may be taken prior to integration. The equation to be constrained is

$$\frac{d^2V}{d\phi_e^2} = -\mu^2 + \frac{3q^4\mu^2}{2\pi^2\lambda} - \frac{\lambda\mu^2}{64\pi^2} + \frac{q^2T^2}{2} + \lambda T^2 f\left(\frac{\mu^2}{T^2}\right) \tag{2.17}$$

where the function $f$ is the second derivative of the thermal bosonic function in [31], evaluated for the scalar boson at $\phi_e = 0$ with the quadratic temperature dependence and the quartic coupling factored out. It is given by

$$f\left(\frac{\mu^2}{T^2}\right) = \frac{1}{2\pi^2} \int_0^\infty dx \frac{x^2 \, e^{-\sqrt{x^2 - \frac{\mu^2}{T^2}}}}{\left(1 - e^{-\sqrt{x^2 - \frac{\mu^2}{T^2}}}\right)\sqrt{x^2 - \frac{\mu^2}{T^2}}}. \tag{2.18}$$

### 2.4.3 Charge $\frac{q}{e} = 1$ Scalar Fields

For a charge equal to the electron charge, satisfying both SSB conditions with $0 < \lambda < 4\pi$ requires the allowed masses of $\phi$ to be too light, much less than the mass of the electron. Such particles would have been produced, for example, at the Large Electron-Positron Collider (LEP) in great quantities and they were not detected, thus ruling out a SSB for $U(1)_{\text{EM}}$ with a $q = e$ field. More concretely, the light masses found here are neatly excluded by the SLAC Anomalous Single Photon (SLAC ASP) search which ruled out $q > 0.08e$ for $m_{\text{MCP}} \lesssim 10$ GeV, a result holding for any weakly interacting millicharged particle [33].

The stability constraint is satisfied for any $\mu^2 < 0$ and any charge $q$ as well as for $\mu^2 > 0$ with ranges of allowed $q$ and $\lambda$, so a heavier electrically charged scalar field with the tree level potential in (2.5) is possible in Nature. However such a scalar could not be the source of spontaneous symmetry breaking for $U(1)_{\text{EM}}$ and could not give the photon a mass.

### 2.4.4 Millicharged (Minicharged) Scalar Fields

Millicharged scalar fields can spontaneously break $U(1)_{\text{EM}}$. Millicharge describes any charge less than that of the electron (i.e. not exclusively $10^{-3}e$), although it may also mean any charge less than that of the down quark, $q < \frac{1}{3}e$. The more accurate (but less used) term is minicharged particles.

The key to a SSB in the finite temperature Universe in this case is to choose parameters that yield an unstable potential at $T = 0$ which gives a SSB in the future and is easily accomplished. The first term in Eqn. (2.17) is negative and dominates the other terms even with a $q = e$ choice for the charge. There are no constraints on the mass in this case, as stated in the previous section.

Next, turn on the finite temperature loop contributions and find parameter ranges that create an overall positive coefficient for the $\phi^2$ terms. The finite temperature pieces are small. The first term arises from the photon running in the loop of the background scalar field. With $T = 2.7K$ or $\approx 10^{-4}$ eV and a millicharge even as large as $q = 10^{-2}e$ it is of order $10^{-12}$. The next term is the scalar boson running in the loop and is bounded by $\lesssim 10^{-8}$ eV for the ranges of masses and charges tested here. The constraint equation becomes

$$\left| -\mu^2 + \frac{3q^4\mu^2}{2\pi^2\lambda} - \frac{\lambda\mu^2}{64\pi^2} \right| \lesssim 10^{-n} \text{ eV}^2 \qquad (2.19)$$

where $10^{-n}$ is defined to be the size of the finite temperature terms. The term inside of the absolute value is negative, its magnitude must be less than that of the positive finite temperature terms. For $\lambda = 1$ and $q = 10^{-3}e$, this gives $m_{\phi_{\text{EM}}} \lesssim 10^{-5}$ eV with a finite temperature term of order $10^{-8}$. Smaller masses and smaller charges are also viable.

Upon inspection of Eqn. 2.17 it appears that small enough values of the quartic coupling $\lambda$ could open up a larger parameter space but in fact this is not so. A very small $\lambda$, $\lambda \ll 1$, can force a positive overall $\phi^2$ coefficient. The condition $\lambda < \frac{3q^2}{2\pi^2}$ forces a stable scalar potential. For $q = 10^{-3}e$, $\lambda < 10^{-12}$ satisfies this with no constraint on the mass of the scalar. On the other hand, the now 2 conditions $\lambda > \frac{3q^4}{2\pi^2}$ and

$$\left| \mu^2 \left( \frac{3q^4}{2\pi^2\lambda} - 1 \right) \right| > \frac{q^2 T^2}{2} \qquad (2.20)$$

force a metastable scalar potential. For the previously considered $q \sim 10^{-3}e$ and $m_{\phi_{\text{EM}}} \sim 10^{-5}$ eV, $\lambda \sim 10^{-13}$ accomplishes the task. The problem is managing a transition between these two states, either by varying $\lambda$ with temperature/time or by some other means. It does not seem possible to do this.

Astrophysical bounds on millicharged particles from stellar cooling constraints set limits on $m \lesssim$ keV masses requiring charges $q \lesssim 10^{-15}e$ [34]. The very light masses considered here allow for charges within these upper bounds. For $q = 10^{-15}e$, $\lambda = 1$, the necessary mass is of the order $m_{\phi_{\text{EM}}} \sim 10^{-5}$ eV. The final term in Eqn. 2.17 is independent of the coupling $q$, thus the scalar mass does not change much from the previous value.

### 2.4.5 $U(1)_{\text{EM}}$ Breaking in the $T > 2.7K$ Universe

It may be that the $U(1)_{\text{EM}}$ has already been broken by a millicharged scalar field. The accepted 2018 limits on the mass of the photon, $m_\gamma \lesssim 10^{-18}$eV, come from magnetohydrodynamic studies of the solar wind [32]. Tighter limits are given by studies of the galactic magnetic field but depend critically on assumptions that may not hold, e.g. the applicability of the virial theorem. However, for the sake of completeness, the tightest limits of $m_\gamma \lesssim 10^{-27}$eV are also discussed here.

The accepted limit of $m_\gamma = 10^{-18}$eV requires a mass for the millicharged field of $m_{\phi_{\text{EM}}} = 3 \times 10^{-13}$ eV for a charge of $q = 10^{-6}e$ and $\lambda = 1$, well within the bounds found here. For $q = 10^{-3}e$ and $q = 10^{-9}e$, the necessary masses are $m_\phi$ of $10^{-16}$ eV and $10^{-10}$ eV respectively.

The smaller $m_\gamma \lesssim 10^{-27}$eV upper bound on the mass of the photon requires $m_\phi = 3 \times 10^{-22}$ with the same charge and quartic coupling value, also within the limits of the model.

Recent work [35] suggests that determining whether the Standard Model photon is exactly massless or not is of interest to light mass dark photon model builders as a strict $m_\gamma = 0$ rules out some of this model parameter space.

In order to accomplish this, it must be that they gain mass via the Stückelberg mechanism ([36] and references within) and not a Higgs mechanism. The SSB mechanism here is able to give photons a mass without putting any restrictions on light dark photon models.

## 2.5 $SU(3)_C$ symmetry breaking - Complementarity

Breaking the symmetry of QCD is not straightforward. The strong coupling $\alpha_s$ is nonperturbative at 2.7 K. The effects of SSB in the 2.7 K Universe require finite temperature field theory calculations. Lattice QCD is needed to calculate the nonperturbative corrections to the effective potential of a new color charged scalar field, $\phi_c^a$, at temperatures below $\Lambda_{QCD}$.

Complementarity between the confined phase of QCD + $\phi_c^a$ and the broken symmetry phase of $SU(3)_C$ may be able to eliminate the need for a SSB to investigate the fate of $SU(3)_C$. A test for the applicability of complementarity has been proposed by Howard Georgi [37]. According to that work, the structure of the heavy stable particles of the confined phase must match that of the broken symmetry phase in order for complementarity to hold.

In order to make use of complementarity, however, a continuously varying parameter must take the confined phase of QCD + a colored scalar field to the spontaneously broken $SU(3)_C$. The mass parameter of a colored scalar field (or fields) is a natural choice as it mirrors the process of $U(1)_{\text{EM}}$. The finite temperature effective potential calculations for colored scalars require the use of lattice QCD. There is a possibility of 3 colored scalar fields being able to manage the transition [38] but the nonperturbative calculations are far beyond the scope of this work. It may be of interest to note that 3 colored scalar fields are 1 more than is necessary for a SSB of $SU(3)_C$ down to no gauge structure at all.

## 2.6 $SU(3)_C$ Symmetry Breaking - SSB

With the applicability of complementarity unclear, the possibility of a carrying out a symmetry breaking of $SU(3)_C$ remains. Both the adjoint and fundamental representations of $SU(3)_C$ are a priori viable as the masses of the gluons are unconstrained in the future. When restricting to SSB, at least 1 colored scalar field in the fundamental representation is added to the theory of QCD and multiple final states of gauge symmetries are possible. For SSB of $SU(3)_C$ to end with no local symmetries, 2 new colored scalar fields are needed. The other possible final symmetry states, a gauged $SU(2)$ or $U(1)$, are realized with one new colored scalar field. An example of a final $SU(2)$ symmetry state is sketched out next.

For $SU(3)_C$ to be spontaneously broken to an $SU(2)$, a colored scalar triplet, $\phi_c^a$ is proposed. This new Higgs field has a $(\mathbf{3}, 0)$ assignment under $SU(3)_C \times U(1)_{\text{EM}}$ and removes one rank from the strong force gauge group, leaving the structure $SU(2)_{BC} \times U(1)_{\text{EM}}$. Here BC stands for "broken color." The minimal effective potential at tree level is

$$V_0(\phi_c) = -m_c^2 \phi_c^\dagger \phi_c + \lambda_c |\phi_c^\dagger \phi_c|^2. \tag{2.21}$$

with color indices suppressed.

On purely dimensional analysis grounds it may be argued that any contribution to $\frac{d^2V}{d\phi_c^2}$ from non-perturbative corrections would be of the order $\Lambda_{QCD}^2$. In order for SSB to occur, it must be that the quantity

$$\left.\frac{d^2V(\phi_c)}{d\phi_c^2}\right|_{\phi_c=0} = -m_c^2 + \Lambda_{QCD}^2 + f_c\left(\frac{m_c^2}{T^2}\right) \tag{2.22}$$

transitions from a positive number to a negative number. The finite temperature pieces $f_c\left(\frac{m_c^2}{T^2}\right)$ are suppressed by the lightest QCD particles, the $\sim 100$ MeV pions. This implies that $m_c^2$ and $\Lambda_{QCD}^2$ are very nearly the same, i.e. the mass of the colored scalar particle would be on the order of hundreds of MeV. Colored scalars in this mass range could bind strongly to single quarks, forming a pion-like system of spin $\frac{1}{2}$ rather than spin $0$. Such mesons would have been detected long ago. In particular, for e⁺e⁻ colliders, the $R$ ratio of the hadronic cross section to the muonic cross section

$$R = \frac{\sigma^{(0)}\left(e^+e^- \to \text{ hadrons }\right)}{\sigma^{(0)}\left(e^+e^- \to \mu^+\mu^-\right)} \tag{2.23}$$

has been extremely well measured [39] and colored scalars of this mass range are ruled out.

A precise calculation is needed for a definitive statement. As mentioned previously, the effects of new colored scalars are highly nontrivial and are not explored further in this thesis. There is a possibility of exploring the idea of new colored scalar fields with light front holographic QCD (LFHQCD) computational tools which have been put to good use for computing nucleon form factors and nucleon spin [40] but this possibility is at the back of the line of projects discussed in Chapter 4.

## 2.7 Conclusions

The possible utility of a new millicharged/minicharged particle has been explored and shown to be a viable candidate for a spontaneous symmetry breaking of the electromagnetic interaction. In addition, the cosmological fate of both the electromagnetic and strong interactions have been investigated under the assumption that we currently live in an intermediate rather than final stage of the symmetries, charges and interactions of particle physics.

It is found that $U(1)_{\text{EM}}$ will remain an infinite range interaction forever for the case of a scalar field with the charge of the electron which is a singlet under $SU(3)_C$.

The more interesting case of a millicharged scalar field, still a singlet under $SU(3)_C$, is viable within a specific range of masses and millicharges. It is capable of spontaneously breaking the symmetry of $U(1)_{\text{EM}}$ and may also be a dark matter candidate [41]. This scenario gives a mass to the photon which is light enough that it would not change the cosmological expansion rate $H(z)$ for $z < 0$ and would likely not affect the fate of the Universe.

$U(1)_{\text{EM}}$ may already have been spontaneously broken by a scalar of mass $m_{\phi_{\text{EM}}} \lesssim 10^{-13}$ eV for a photon mass equal to the current upper bound of $m_\gamma \sim 10^{-18}$ eV and a charge of $q = 10^{-6}e$. If the astrophysical limits from the galactic magnetic field studies on $m_\gamma$ hold, this upper bound becomes $m_\gamma \sim 10^{-27}$ and the mass of the millicharged field $m_{\phi_{\text{EM}}} \lesssim 10^{-22}$ again for microcharges $q = 10^{-6}e$. The higher temperatures of the earlier Universe appear to lift the low mass constraint on $m_\phi$ and this will be explored in followup work [41]. This would push $\phi_{\text{EM}}$ into the realm of detection with the dark matter experiments discussed in Section 2.2.

The fate of $SU(3)_C$ is likely to remain unbroken but is as yet unknown.

If it emerges that $U(1)_{\text{EM}}$ was broken at higher temperatures in the early Universe or will be broken in the future, there may be some aesthetic appeal to either breaking the one remaining symmetry of $SU(3)_C$ or forbidding it from being broken. Should this be the case, future work would include testing Georgi's complementarity principle in earnest with collaborators in the QCD community. The successful implementation of complementarity along with a broken $U(1)_{\text{EM}}$ would yield a final state of the Universe with no local symmetries at all, satisfying an evolution from early Universe higher symmetry structures to none at all at late times.

This concludes the past research effort into a minimal, one scalar field, extension of known symmetries of Nature.

The thread of this chapter will be picked up again in Chapter 4 as it leads directly to a current project properly examining the constraints upon $\phi_{\text{EM}}$ if it is to be considered a dark matter candidate. There is a resurgence of research interest into millicharged/minicharged particles - both scalar and fermionic - for near term and futuristic direct detection experiments [42]. Previous experiments such as those carried out at SLAC National Accelerator Laboratory put constraints on new minicharged particles, e.g. the SLAC Linear Collider (SLC) which wrapped up suddenly in 1998 [43]. The SLC was an electron positron collider and this particular experiment was designed to discover whether electric charge is quantized or not.

It seemed (and still seems) evident that Nature quantizes charge in units of the magnitude of the down quark charge, $q = \frac{e}{3}$, yet the Standard Model describing all known particle physics forces does not mandate charge quantization. As discussed in Chapter 1, physical phenomena that are not mandatory by theory are usually forbidden. Discovery of a particle with charge less than that of the down quark, provided it does not itself become the new quanta of electric charge, would force the Standard Model into the "charge quantization forbidding" regime [44]. This greatly assists physicists building models to expand the known laws of physics in order to solve the outstanding problems in particle physics and cosmology.

# Chapter 3

# Dark Matter and Baryogenesis from a New Local Symmetry

## 3.1 Overview

The research of this chapter enters the vast arena of possible new symmetries of Nature. A model of particle dark matter that solves more than one of the outstanding mysteries of particle physics today and relies upon a new gauge symmetry is proposed. In this model, the Standard Model gauge symmetry is extended by a new $SU(2)_\ell$ group acting nontrivially on only the lepton sector (as indicated by the subscript $\ell$). This new symmetry is spontaneously broken at the TeV scale, well above the electroweak symmetry breaking scale of $\sim 100$ GeV. Under the new $SU(2)_\ell$ ordinary leptons form doublets along with new lepton partner fields. This construction naturally contains a dark matter candidate, the partner of the right-handed neutrino, stabilized by a residual global $U(1)_\chi$ symmetry. The model contains baryogenesis through an asymmetric dark matter scenario in which generation of related asymmetries in the dark matter and baryon sectors are driven by the $SU(2)_\ell$ instantons during a first order phase transition in the early Universe.

## 3.2 Introduction

Gravitational evidence for the existence of dark matter composing $\sim 26\%$ of the energy density of the Universe is extremely compelling [8]. As outlined in Chapter 1, it includes measurements of galactic rotation curves, gravitational lensing, cosmic microwave background anisotropy, $X$-ray emission from elliptical galaxies and collisions of clusters of galaxies. The Standard Model of particle physics in its current form does not account for the presence of dark matter. However, if dark matter is a particle or set of particles that couples appreciably to visible matter, it may be reasonable to expect that it represents an extension of the SM built on similar principles, i.e. those of a local symmetry (gauge) theory.

One possible path to model dark matter is to postulate a larger gauge symmetry group, e.g. a new $SU(N)$ or $U(N)$ that is a direct product with the known $SU(3)_\text{C} \times SU(2)_\text{W} \times U(1)_\text{Y}$ group structure. Another possibility is to unify the known gauge symmetries into a higher rank group, e.g. an $SU(5)$ or $SO(10)$. In this chapter, the first path is taken. Any proposed new symmetry structure should play a primary role in explaining why dark matter is stable on the scale of the age of the Universe and how it is forbidden from decaying into the known Standard Model particles.

In addition to the mystery of dark matter, the SM is unable to explain the observed matter-antimatter asymmetry of the Universe. Theories proposed to solve this problem are known as baryogenesis models rather than matter-o-genesis models due to the uncertainty in cosmological abundances of leptons vs. antileptons, most notably in the neutrino sector. In 1967 the great Russian physicist Andrei Sakharov proposed three conditions that any theory of baryogenesis must satisfy [46]: violation of baryon number, violation of charge conjugation $C$ as well as the product of $C$ and the parity operator $P$, and out of equilibrium dynamics. Two of the three Sakharov conditions, the requirement of a first order phase transition and sufficient amount of $CP$ violation, appear not be fulfilled in the SM. There are many ideas for how to fix this by introducing new fields and interactions as a way to go Beyond the Standard Model. Similarly to the approach used for the dark matter problem, one class of potential solutions relies on extensions of the Standard Model gauge group. For example, it was shown [47] that the breakdown of a new gauge group can overcome the SM difficulties and provide a framework for successful baryogenesis.

Physical theories that unify and combine not only symmetries but other mechanisms as well can further our understanding of the natural world in powerful ways. The theory put forth in this chapter uses a combination of ideas to suggest a common solution to the identity of dark matter and the matter/antimatter asymmetry, namely by stating that DM interactions with the SM are responsible both for generating its abundance as well as generating the observed baryon asymmetry. This is the philosophy behind asymmetric dark matter (ADM) models [48, 49, 50, 51, 52, 53], in which the DM and baryon asymmetries are intimately related and in fact arise from the same underlying mechanism. A typical feature of these models is that the natural scale for the DM mass is $\sim \text{GeV}$, which requires light messengers in order to realize a large enough DM annihilation cross section to annihilate away the symmetric component. The research in this chapter constructs an ADM model that naturally contains such particles.

Beginning as usual with symmetries, gauged extensions of the SM global symmetries are considered. The possibility of gauging lepton and baryon number was considered in [54, 55, 56, 57, 58, 59]. Unfortunately, models based on this approach offer only a limited possibility of explaining the primordial baryon asymmetry [60, 61, 62]. More recently, a baryogenesis mechanism based on a non-Abelian extension of the SM baryon number and color in $SU(4)$ gauge group was proposed in [63, 64]. Such an extension successfully unifies DM with the SM baryons, but ultimately relies on unspecified UV physics represented by higher-dimensional operators to generate the asymmetry. As mentioned at the end of this chapter, it may well be that the choice of extending the interactions of Nature based on conservation laws that hold only at the classical level and break down in the quantum regime limits the power of any theory built upon them. Chapter 4 explores this further.

This chapter follows the general approach of [64] except in the lepton sector rather than the baryonic section. It extends the SM gauge group by an additional $SU(2)_\ell$ gauge symmetry under which the SM leptons transform nontrivially, promoting them to $SU(2)_\ell$ doublets along with additional partner fields. A lepton number assignment is extended to the partner fields, thus generalizing the SM lepton number as must be the case for an anomalous global symmetry which cannot be gauged itself. The $SU(2)_\ell$ symmetry is spontaneously broken via a vacuum expectation value (vev) of a new leptonic Higgs field $\Phi$. All new matter fields introduced in the model obtain vector-like masses after $SU(2)_\ell$ breaking. The lightest of these particles will be stable due to residual global symmetries and provides a dark matter candidate of ADM framework. During the $SU(2)_\ell$ phase transition, $SU(2)_\ell$ sphalerons generate both DM and lepton number, with the latter later converted to baryon number by electroweak sphalerons. A similar use of sphalerons is described in the aidnogenesis scenario [65].

## 3.3 Beyond the Standard Model extension

The gauge symmetry of the Standard Model is extended by a single non-Abelian gauge symmetry:

$$SU(3)_c \times SU(2)_W \times U(1)_Y \times SU(2)_\ell \ . \tag{3.1}$$

The SM leptons reside in upper components of $SU(2)_\ell$ doublets, while the new fermions (denoted by a tilde) reside in lower components of $SU(2)_\ell$ doublets:

$$\hat{l}_L \equiv \begin{pmatrix} L \\ \tilde{l}_L \end{pmatrix}, \quad \hat{e}_R \equiv \begin{pmatrix} e_R \\ \tilde{e}_R \end{pmatrix}, \quad \hat{\nu}_R \equiv \begin{pmatrix} \nu_R \\ \tilde{\nu}_R \end{pmatrix}. \qquad (3.2)$$

To maintain the Standard Model cancellation of $SU(2)_W \times U(1)_Y$ anomalies and allow new fermions to acquire Dirac masses after $SU(2)_\ell$ symmetry breaking a set of leptons that are neutral under $SU(2)_\ell$ is introduced,

$$l'_R, \quad e'_L, \quad \nu'_L. \qquad (3.3)$$

This is a minimal and compulsory addition to the theory in order for it to respect gauge invariance. After $SU(2)_\ell$ is broken, these fields combine with $\tilde{l}_L$, $\tilde{e}_R$, and $\tilde{\nu}_R$ to form massive Dirac fermions.

Finally, two Higgs doublets charged under $SU(2)_\ell$ are introduced. The second doublet is necessary to provide two of the required ingredients for a realistic ADM model. First, it will supply sufficient CP violation to catalyze the production of an adequate baryon asymmetry. Second, it will provide an annihilation channel that allows for sufficient annihilation of the symmetric component of the DM. These are required to match cosmological observations.

The quantum numbers for all the relevant particles including two $SU(2)_\ell$ Higgs doublets (discussed below) are summarized in Table 3.1.

The new local symmetry must be broken for many reasons, primarily that it is the breaking process that begets stable dark matter and baryogenesis. However, it is also the case that there are strict limits on infinite range forces which arise from unbroken gauge groups [66, 67] and this is to be avoided.

| Field | $SU(2)_\ell$ | $SU(2)_W$ | $U(1)_Y$ |
|---|---|---|---|
| $\hat{l}_L = \begin{pmatrix} l_L \\ \tilde{l}_L \end{pmatrix}$ | 2 | 2 | $-1/2$ |
| $\hat{e}_R = \begin{pmatrix} e_R \\ \tilde{e}_R \end{pmatrix}$ | 2 | 1 | $-1$ |
| $\hat{\nu}_R = \begin{pmatrix} \nu_R \\ \tilde{\nu}_R \end{pmatrix}$ | 2 | 1 | 0 |
| $l'_R$ | 1 | 2 | $-1/2$ |
| $e'_L$ | 1 | 1 | $-1$ |
| $\nu'_L$ | 1 | 1 | 0 |
| $\Phi_1, \Phi_2$ | 2 | 1 | 0 |

Table 3.1: Fields and their representations under the gauge symmetries $SU(2)_\ell \times SU(2)_W \times U(1)_Y$.

In order to spontaneously break $SU(2)_\ell$, the following scalar potential for $SU(2)_\ell$ doublets $\Phi_i$ is introduced:

$$\begin{aligned} V(\Phi_1, \Phi_2) &= m_1^2|\Phi_1|^2 + m_2^2|\Phi_2|^2 + (m_{12}^2 \Phi_1^\dagger \Phi_2 + \text{h.c.}) + \lambda_1|\Phi_1|^4 + \lambda_2|\Phi_2|^4 + \lambda_3|\Phi_1|^2|\Phi_2|^2 \\ &+ \lambda_4|\Phi_1^\dagger \Phi_2|^2 + \left[\tilde{\lambda}_5 \Phi_1^\dagger \Phi_2 |\Phi_1|^2 + \tilde{\lambda}_6 \Phi_1^\dagger \Phi_2 |\Phi_2|^2 + \tilde{\lambda}_7 (\Phi_1^\dagger \Phi_2)^2 + \text{h.c.}\right] \end{aligned} \quad (3.4)$$

where a $U(1)_1$ symmetry (discussed below) is imposed to ensure that the dark matter is ultimately stable. In addition, while this research did not explicitly show $\Phi_i$ interactions with the SM Higgs, such interactions are generically present at tree level and are induced radiatively even if not present. This interaction plays an important role in allowing the lightest component of $\Phi$ to decay through induced mixing with the SM Higgs, with constraints discussed in Sec. 3.5.

The potential contains four complex parameters: $m_{12}^2$, $\tilde{\lambda}_5$, $\tilde{\lambda}_6$, and $\tilde{\lambda}_7$. For generic parameters, one phase can be rotated away by redefining the phase of the combination $\Phi_1^\dagger \Phi_2$ (the only combination appearing in the potential), leaving three physical phase combinations [68].

It is straightforward to choose parameters so that $SU(2)_\ell$ is completely broken by $\Phi_{1,2}$ vevs. The potential, Eq. (3.4), is structurally identical to a two Higgs doublet model (with the global $U(1)_1$ playing the role of the SM's gauged $U(1)_Y$ hypercharge) and admits a similarly rich array of mass eigenstates for the physical bosons. The vacuum can be parameterized by $v_\ell = \sqrt{v_1^2 + v_2^2}$ and $\tan\beta = v_1/v_2$, where $v_1$ and $v_2$ are the vevs of the two doublets, respectively, $\langle \Phi_1 \rangle = \frac{1}{\sqrt{2}}(0, v_1)^T$ and $\langle \Phi_2 \rangle = \frac{1}{\sqrt{2}}(0, v_2)^T$. There is a spectrum of five physical scalar Higgs bosons which are mixtures of the original CP-even and CP-odd components of the doublets $\Phi_1$ and $\Phi_2$.

The Yukawa interactions consistent with the gauge symmetries are given by

$$\begin{aligned}
\mathscr{L}_Y &= \sum_i \left( Y_l^{ab} \, \bar{\hat{l}}_L^a \, \Phi_i \, l_R'^b + Y_e^{ab} \, \bar{\hat{e}}_R^a \, \Phi_i \, e_L'^b + Y_\nu^{ab} \, \bar{\hat{\nu}}_R^a \, \Phi_i \, \nu_L'^b \right) \\
&+ y_e^{ab} \, \bar{\hat{l}}_L^a \, H \, \hat{e}_R^b + y_\nu^{ab} \, \bar{\hat{l}}_L^a \, \tilde{H} \, \hat{\nu}_R^b + y_e'^{ab} \, \bar{l}_R^a \, H \, e_L'^b + y_\nu'^{ab} \, \bar{l}_R^a \, \tilde{H} \, \nu_L'^b + \text{h.c.}
\end{aligned} \quad (3.5)$$

where $a$ and $b$ are flavor indices.

After electroweak symmetry breaking the Yukawa matrices $y_e^{ab}$ and $y_\nu^{ab}$ lead to the usual lepton mass matrices, now with Dirac neutrino masses. The newly introduced Yukawa matrices $Y_l$, $Y_e$, $Y_\nu$, $y_e'$ are responsible for generation of Dirac mass terms between the lepton partners and spectators after $SU(2)_\ell$ breaking.

The full mass matrix for new fields includes mixing between electroweak singlets and doublets induced by the SM Higgs vev:

$$\frac{1}{\sqrt{2}} \left( \bar{\tilde{\nu}}_L \; \bar{\nu}'_L \right) \begin{pmatrix} Y_l v_\ell & y_\nu v \\ y'^\dagger_\nu v & Y^\dagger_\nu v_\ell \end{pmatrix} \begin{pmatrix} \nu'_R \\ \tilde{\nu}_R \end{pmatrix} + \frac{1}{\sqrt{2}} \left( \bar{\tilde{e}}_L \; \bar{e}'_L \right) \begin{pmatrix} Y_l v_\ell & y_e v \\ y'^\dagger_e v & Y^\dagger_e v_\ell \end{pmatrix} \begin{pmatrix} e'_R \\ \tilde{e}_R \end{pmatrix} + \text{h.c.} \quad (3.6)$$

where each field should be understood as a three-component vector in flavor space, with the Yukawa couplings $3 \times 3$ complex matrices, and the vevs $v$ and $v_\ell$ belong to the SM Higgs and the new Higgses, respectively.

The couplings $y_e$ and $y_\nu$ are required to reproduce the SM flavor structure and are therefore tiny. If in addition $y'_{\nu,e} v \ll Y_{\ell,\nu,e} v_\ell$, there is little mixing between doublets and singlets, and negligible contributions to precision electroweak observables. It is both possible and highly convenient to choose parameters in this regime and this is what is done. It would be ideal to be forced into this parameter space explicitly by the theory but that is not the case here.

## 3.4 Baryogenesis prior to electroweak symmetry breaking

This section describes the $SU(2)_\ell$ phase transition which provides the third and final Sakharov criteria, namely the out of equilibrium condition necessary to evolve a nonzero baryon number.

### 3.4.1 Nonperturbative dynamics

Nonperturbative dynamics lead to simultaneous leptogenesis and dark matter genesis through sphaleron processes during the $SU(2)_\ell$ phase transition. To understand the generation of lepton and DM numbers during the $SU(2)_\ell$ transition an analysis of the global symmetries of the model is required. The analysis is complicated as it is due to the choice of a leptonic symmetry for the SM extension. Lepton number is a classical symmetry, not a quantum symmetry, i.e. it is anomalous and cannot be gauged. The conserved quantity in this chapter is a generalization of lepton number and its subgroups are intuitively obtained as they would be for a non-anomalous $U(1)_\text{L}$.

|  | Symmetries ||||||||| 
|  | Exact & Approximate ||| Lepton Basis ||| Low Energy |||
| Field | $U(1)_1$ | $U(1)_2$ | $U(1)'$ | $U(1)_1$ | $U(1)_L$ | $U(1)_\chi$ | $U(1)_D$ | $U(1)_L$ | $U(1)_\chi$ |
| --- | --- | --- | --- | --- | --- | --- | --- | --- | --- |
| $\hat{l}_L = \begin{pmatrix} l_L \\ \tilde{l}_L \end{pmatrix}$ | 0 | 1 | 1 | 0 | 1 | 0 | $\begin{matrix}-1\\1\end{matrix}$ | 1 | 0 |
| $\hat{e}_R = \begin{pmatrix} e_R \\ \tilde{e}_R \end{pmatrix}$ | 0 | 1 | 1 | 0 | 1 | 0 | $\begin{matrix}-1\\1\end{matrix}$ | 1 | 0 |
| $\hat{\nu}_R = \begin{pmatrix} \nu_R \\ \tilde{\nu}_R \end{pmatrix}$ | 0 | 1 | $-1$ | 0 | 0 | 1 | $\begin{matrix}-1\\1\end{matrix}$ | 0 | 1 |
| $l'_R$ | 1 | 1 | 1 | 1 | 1 | 0 | 1 | 1 | 0 |
| $e'_L$ | 1 | 1 | 1 | 1 | 1 | 0 | 1 | 1 | 0 |
| $\nu'_L$ | 1 | 1 | $-1$ | 1 | 0 | 1 | 1 | 0 | 1 |
| $\Phi_i,\ i=1,2$ | $-1$ | 0 | 0 | $-1$ | 0 | 0 | $\begin{matrix}-2\\0\end{matrix}$ | 0 | 0 |

Figure 3.1: Charges under global $U(1)$ symmetries. The first two columns represent charges under exact global symmetries of the Lagrangian. The next four columns represent charges under exact and approximate global symmetries when $y_\nu$ and $y'_\nu$ can be neglected. The last three columns represent charges under exact and approximate global symmetries in low energy physics.

The theory possesses two anomaly-free global $U(1)$ symmetries consistent with the gauge structure and Yukawa interactions in Eq. (3.5) (see the first panel of Figure 3.1). In a realistic model neutrino Yukawa couplings $y_\nu$ and $y'_\nu$ must be small, therefore an additional approximate global symmetry $U(1)'$ exists. The $U(1)'$ global symmetry is anomalous under $SU(2)_\ell$ and thus will be broken by instanton generated interactions. For the purpose of leptogenesis it is convenient to construct linear combinations of the $U(1)_2$ and $U(1)'$ symmetries that will correspond to generalized lepton and DM numbers. Charge assignments under these symmetries, $U(1)_L$ and $U(1)_\chi$ respectively, are shown in the middle panel of Figure 3.1.

The $U(1)_1$ symmetry is spontaneously broken by the $\Phi_1$ and $\Phi_2$ vevs. However, a global $U(1)_D$

subgroup of $SU(2)_\ell \times U(1)_1$ survives. This unbroken $U(1)_D$ is a diagonal combination of the $U(1)_1$ and the $U(1)$ group generated by the $\tau_3$ generator of $SU(2)_\ell$. The charges of the fields in the low energy theory are shown in the last panel of Figure 3.1. Note that the charges of light fields under $U(1)_D$ are given by the sum of lepton and DM charges and thus $U(1)_D$ is not visible in low energy physics. However, $U(1)_D$ distinguishes between the SM leptons and new particles and thus will be responsible for the stability of the DM.

Both $U(1)_L$ and $U(1)_\chi$ are individually anomalous under $SU(2)_\ell$ interactions. On the other hand, the sum of lepton and DM numbers will be conserved since it corresponds to an anomaly free $U(1)_2$ symmetry. This means that $SU(2)_\ell$ instantons generate effective interactions in low energy theories that break $U(1)_L$ and $U(1)_\chi$ individually while conserving the sum of the two charges, $L + \chi$. For illustrative purposes, it is convenient to consider a one flavor (generation) toy model. The $SU(2)_\ell$ instantons generate an effective 4-fermion interaction that involves all of the doublets of $SU(2)_\ell$. Applying results from [69], the single generation sphalerons can be represented as a dimension six operator,

$$\mathcal{O}_{\text{eff}} \sim \epsilon_{ij} \left[ (l_L^i \cdot \bar{\nu}_R)(l_L^j \cdot \bar{e}_R) - (l_L^i \cdot \bar{\nu}_R)(\tilde{l}_L^j \cdot \bar{\tilde{e}}_R) + (l_L^i \cdot \tilde{l}_L^j)(\bar{\nu}_R \cdot \bar{\tilde{e}}_R) \right.$$
$$\left. - (l_L^i \cdot \tilde{l}_L^j)(\bar{\tilde{\nu}}_R \cdot \bar{e}_R) + (\tilde{l}_L^i \cdot \bar{\nu}_R)(\tilde{l}_L^j \cdot \bar{\tilde{e}}_R) - (\tilde{l}_L^i \cdot \bar{\tilde{\nu}}_R)(l_L^j \cdot \bar{e}_R) \right], \quad (3.7)$$

where the dots denote Lorentz contractions, $i, j$ are $SU(2)_W$ indices and the $SU(2)_\ell$ indices have been expanded out.

Upon inspection of the instanton induced interactions in Eq. (3.7), violation of lepton and DM numbers by $\Delta L = -1$ and $\Delta \chi = 1$ is apparent. For example, consider only the last term in Eq. (3.7). It is responsible for processes $\nu_L \tilde{e}_L \to \tilde{\nu}_R e_R$ and $\tilde{\nu}_L e_L \to \tilde{\nu}_R e_R$. Since $l_L, \tilde{l}_L, e_R$ and $\tilde{e}_R$ have $L = 1$ while $\nu_R$ and $\tilde{\nu}_R$ have $\chi = 1$, the instanton-induced interactions violate lepton and DM numbers by $\Delta L = -1$ and $\Delta \chi = 1$.

Generalization to a three generation model is straightforward and leads to a 12 fermion operator. At zero temperature the instanton operator is exponentially suppressed but at high temperatures the $SU(2)_\ell$ symmetry is restored and lepton- and DM-number violating interactions are unsuppressed. The combined effect of all instanton-induced interactions (3.7) is calculated numerically by solving the particle diffusion equations (see Sec. 3.4.2).

It might appear that the lepton and DM numbers would be immediately washed out since both $U(1)_L$ and $U(1)_\chi$ are explicitly broken by Yukawa interactions. However, the right-handed neutrinos and their partners reach chemical equilibrium long after the $SU(2)_\ell$ phase transition because of the smallness of their Yukawa couplings. In the case of the SM neutrinos, this is implied by the small observed neutrino masses. In the case of the neutrino partners, this requires the interaction rate $\Gamma(H \leftrightarrow \ell'\nu')$ to be much less than the Hubble expansion rate at the $SU(2)_\ell$ phase transition, which will be satisfied provided $y'_\nu \lesssim 10^{-6}$ for $u \sim$ TeV. As a result, both the lepton and DM number asymmetries survive until the electroweak transition. Thus during and just prior to electroweak symmetry breaking the electroweak sphalerons are able to act upon the lepton number deficit and transfer it into baryons through the Dirac leptogenesis mechanism [70, 71].

### 3.4.2 Phase transition

In order to investigate the phase transition from the false to true vacuum states an analysis of the scalar potential is required.

**Finite temperature effective potential**

The one-loop effective scalar potential at nonzero temperature can be written in terms of the background field $u$ as

$$V(u,T) = V_{\text{tree}}(u) + V_{1\,\text{loop}}(u,0) + V_{\text{temp}}(u,T) \,, \tag{3.8}$$

where the first of the individual contributions is the tree-level part,

$$V_{\text{tree}}(u) = -\frac{1}{2}m^2\,u^2 + \frac{1}{4}\lambda\,u^4 \,, \tag{3.9}$$

and the mass parameter $m^2$ and quartic $\lambda$ schematically indicate combinations of those parameters from the scalar potential, Eq. (3.4). The remaining terms on the right-hand side correspond to the zero temperature Coleman-Weinberg correction and the one-loop finite temperature contribution.

To calculate the Coleman-Weinberg term, the cut-off regularization scheme is implemented and it is assumed that the minimum of the one-loop potential and the $SU(2)_\ell$ Higgs mass are the same as their tree-level values (see, e.g. [31]) and the previous chapter. The zero-temperature one-loop correction takes the form,

$$V_{1\,\text{loop}}(u) = \frac{1}{64\pi^2} \sum_i n_i \left\{ m_i^4(u) \left[ \log\left(\frac{m_i^2(u)}{m_i^2(v_\ell)}\right) - \frac{3}{2} \right] + 2\,m_i^2(u)\,m_i^2(v_\ell) \right\} \,, \tag{3.10}$$

where the sum is over all particles charged under $SU(2)_\ell$ and $n_i$ denoted the number of degrees of freedom, with an extra minus sign for the fermions.

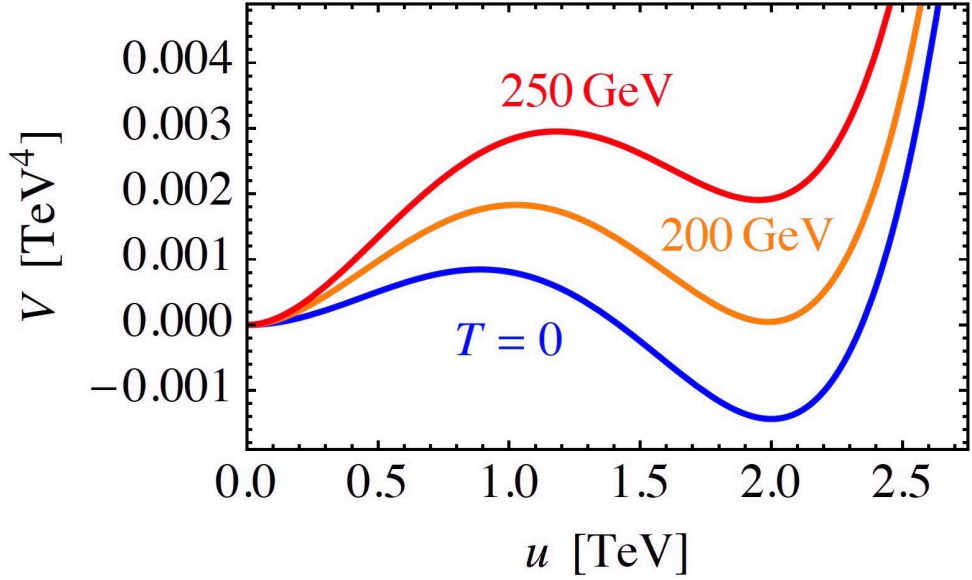

Figure 3.2: Plot of the finite temperature effective potential $V(u, T_c)$ for $v_\ell = 2$ TeV, $\lambda_1 = 2 \times 10^{-3}$, $g_\ell = 1$ and $T_c = 200$ GeV.

Using the formula for the one-loop finite temperature correction [31], the temperature dependent piece is

$$V_{\rm T}(u, T) = \frac{T^4}{4\pi^2} \sum_i n_i (3 \mp 1) \int_0^\infty dx\, x^2 \left[ \log\left(1 \mp e^{-\sqrt{x^2 + m_i^2(u)/T^2}}\right) - \log\left(1 \mp e^{-x}\right) \right]. \quad (3.11)$$

In the expression above the sum is again over all fields, $n_i$ is the number of degrees of freedom that contains a factor of $-1$ for fermions as just mentioned. Inside the $\pm$ symbol, minus signs are for bosons and the plus signs are for fermions.

At this point, the shape of the full effective potential for various temperatures may be analyzed. For successful baryogenesis the phase transition has to be strongly first order, $v_\ell(T_c)/T_c \gtrsim 1$, which favors small values of the effective quartic $\lambda$. The constraint on $v_\ell$ coming from the LEP-II experiment is $v_\ell \gtrsim 1.7$ TeV [62]. Finally, the critical temperature of the phase transition should not be lower than $\sim 175$ GeV in order to occur before the electroweak phase transition. Figure 3.2 shows the plot of $V(u, T_c)$ for sample parameter values fulfilling those constraints: $g_\ell = 1, v_\ell = 2$ TeV and $\lambda = 2 \times 10^{-3}$, leading to $T_c \sim 200$ GeV.

**Bubble nucleation and diffusion equations**

A first order phase transition, if possible, takes place at the critical temperature $T_c$. Bubbles of true vacuum are nucleated and then expand, eventually colliding and filling the entire universe. Following [47] and denoting the bubble radius by $R$ and the width of the wall of the bubble by $L_w$, the following ansatz for the bubble profile,

$$u(r) = \frac{1}{2} u_c \left[ 1 - \tanh\left(\frac{r - R}{L_w}\right) \right] \tag{3.12}$$

is considered. In this case the width of the bubble scales as $L_w \sim 1/T$. The expanding bubble is assumed to be large, so that to a good approximation its evolution is in one dimension, along the $z$ axis, which is taken to be perpendicular to the bubble wall. The bubble wall is placed at $z = 0$ with the broken phase on the $z > 0$ side and the unbroken phase into which the bubble expands is on the $z < 0$ side. A bubble wall velocity of $v_w \approx 0.05 \, c$ is adopted, although I note that values ranging from $0.1 \, c$ to $c/\sqrt{3} \simeq 0.6 \, c$ are given as plausible in the excellent early work on diffusion and transport in electroweak baryogenesis [72]. It will turn out that this choice does not greatly affect the ability to produce a viable lepton number asymmetry (3.25).

In the presence of CP violation, the $SU(2)_\ell$ instantons produce lepton and DM number asymmetries. In order to estimate their magnitude, a set of coupled diffusion equations is solved for particle number densities [72, 47]. Since only leptons are affected by the presence of the new gauge group, there are 12 relevant equations in the work presented here (see Appendix A for diffusion equations and constraints). The 12 equations involve the following particle number densities:

$$n(l) = n(e_L) + n(\nu_L)\,, \;\; n(e) = n(e_R),\;\; n(\nu) = n(\nu_R)\,,\;\; n(\tilde{l}) = n(\tilde{e}_L) + n(\tilde{\nu}_L) \quad (3.13)$$

$$n(\tilde{e}) = n(\tilde{e}_R)\,,\;\; n(\tilde{\nu}) = n(\tilde{\nu}_R)\,,\;\; n(l') = n(e'_R) + n(\nu'_R)\,,\;\; n(e') = n(e'_L) \quad\quad (3.14)$$

$$n(\nu') = n(\nu'_L)\,,\;\; n(h) = n(h^+) + n(h^0)\,,\;\; n(\Phi^u) = n(\Phi^u_1)\,,\;\; n(\Phi^d) = n(\Phi^d_2)\,. \quad (3.15)$$

There are nine constraints on the particle number densities corresponding to the Yukawa equilibrium conditions as well as four constraints coming from the instanton equilibrium requirement (see Appendix A.1). However, not all of these constraints are independent. Only seven of the Yukawa equilibrium conditions and one of the instanton equilibrium conditions are linearly independent.

The diffusion equations contain diffusion constants for each particle species. For particles charged under the SM, their magnitude has been estimated in Ref. [73]. Carrying out a similar calculation and taking the $SU(2)_\ell$ gauge coupling to be $g_\ell \approx 1$ the following estimates are obtained,

$$D_l = D_{\tilde{l}} \sim D_e = D_{\tilde{e}} \sim D_\nu = D_{\tilde{\nu}} \sim D_{\Phi^u} = D_{\Phi^d} \sim 25/T\,,\;\; D_h = D_{l'} \sim \frac{100}{T}\,,$$

$$D_{e'} \sim D_{\nu'} \sim \frac{400}{T}\,. \quad (3.16)$$

This process yields four coupled equations for the $l, \tilde{\nu}, \phi^u$ and $h$ particle number densities

$$v_w \left[4n'(l) + 4n'(\tilde{\nu}) - 2n'(\Phi^u) - n'(h)\right] -$$
$$\frac{25}{T} \left[22\, n''(l) + 4\, n''(\tilde{\nu}) - 20\, n''(\Phi^u) - n''(h)\right] = 0 , \qquad (3.17)$$

$$v_w \left[2n'(l) + n'(\Phi^u)\right] - \frac{25}{T} \left[2\, n''(l) + n''(\Phi^u)\right] = \gamma_1\, \theta(L_w - |z|) , \qquad (3.18)$$

$$v_w \left[-n'(l) + 6\, n'(\tilde{\nu}) + n'(\Phi^u) - \tfrac{3}{2}\, n'(h)\right] -$$
$$\frac{25}{T} \left[-n''(l) + 6\, n''(\tilde{\nu}) + n''(\Phi^u) - \tfrac{3}{2}\, n''(h)\right] = \gamma_2\, \theta(L_w - |z|) , \qquad (3.19)$$

$$v_w \left[\tfrac{5}{2}\, n'(h)\right] - \frac{25}{T} \left[13\, n''(h)\right] = 0 , \qquad (3.20)$$

where the primes denote derivatives with respect to $z$ and $\gamma_1, \gamma_2$ are the CP-violating sources for the new $SU(2)_\ell$ Higgs induced by the bubble wall.

The values for $\gamma_1$ and $\gamma_2$ in the two Higgs doublet model have been derived in [74] and are given by

$$\gamma_i(z) \approx \frac{\tilde{\lambda}_7}{32\pi} \Gamma_{\phi_i} T \frac{m_{12}^2}{m_{\phi_i}^3(T)} \partial_{t_z} \phi_i, \qquad (3.21)$$

where $\tilde{\lambda}_7$ and $m_{12}^2$ are free parameters from the scalar potential.

One can choose parameter values such that $\gamma_1 \approx \gamma_2 \approx 5 \times 10^{-5}$ GeV$^4$, which yields a baryon to entropy ratio of $\sim 10^{-10}$ and is quite close to the observed ratio [8].

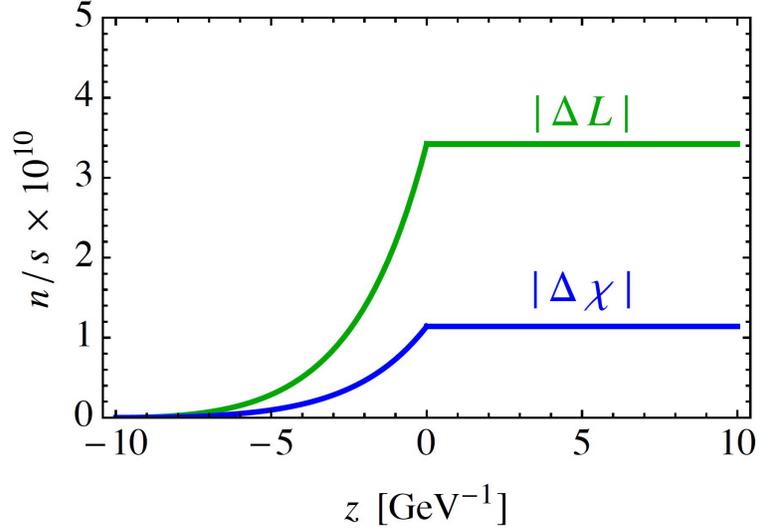

Figure 3.3: SM lepton and DM particle number densities as a function of the spatial position $z$ assuming a bubble wall located at $z = 0$.

**Lepton and dark matter asymmetries**

The particle number densities corresponding to the globally conserved $U(1)$ charges given in Figure 3.1

$$\begin{aligned} \Delta L(z) &= n(l) + n(\tilde{l}) + n(l') + n(e) + n(\tilde{e}) + n(e'), \\ \Delta \chi(z) &= n(\nu) + n(\tilde{\nu}) + n(\nu'), \end{aligned} \qquad (3.22)$$

are the number densities of interest for leptogenesis.

Use of the equilibrium conditions yields

$$\Delta L(z) = 3\left[n(l) + n(\tilde{\nu}) - \frac{1}{2}n(\Phi^u) - \frac{1}{2}n(h)\right], \qquad (3.23)$$

and

$$\Delta \chi(z) = \left[ n(l) + n(\tilde{\nu}) - \frac{1}{2} n(\Phi^u) + \frac{1}{2} n(h) \right] . \tag{3.24}$$

Figure 3.3 shows the solutions of the diffusion equations assuming $T_c = 200$ GeV and $\gamma_1 = \gamma_2 = 5 \times 10^{-5}$ GeV$^4$.

The ratio of the produced lepton and DM asymmetries is

$$\left| \frac{\Delta L}{\Delta \chi} \right| = 3 , \tag{3.25}$$

and is roughly independent of the numerical values of $v_w$, $T_c$, $\gamma_1$, and $\gamma_2$. This perhaps justifies the choice of values which was somewhat arbitrary, certainly in the case of $v_w$.

**Baryon asymmetry for baryogenesis**

The particle number densities in Fig. 3.3 are normalized to the entropy $s \approx (2\pi^2/45) g_* T^3$, with the number of relativistic degrees of freedom $g_* \sim 100$. For this set of parameters, the ratio of the lepton number density and the entropy is roughly $\Delta L/s \sim 3 \times 10^{-10}$. The lepton asymmetry is $n_L/s \approx 3 \times 10^{-10}$.

The $SU(2)_\ell$ instanton induced interactions shut off after the $SU(2)_\ell$ breaking concludes and the DM asymmetry freezes in. However, above the electroweak phase transition the SM sphalerons are active and they convert part of the SM lepton asymmetry to a baryon asymmetry. The baryon asymmetry generated by the sphalerons is [75],

$$\Delta B = \frac{28}{79}\Delta L \ . \tag{3.26}$$

This result depends slightly on the lepton partner masses. The dependence is minimized if those masses are below the electroweak scale, $T \sim 100$ GeV. The final generated baryon asymmetry to entropy ratio is therefore,

$$\frac{n_B}{s} \approx 10^{-10} \ , \tag{3.27}$$

which is the same order of magnitude as cosmological observations.

## 3.5 Dark matter

The DM is a mixed state, largely composed of the lightest $\tilde{\nu}_R$. Through interactions with the SM Higgs, it picks up a small component of the electroweak doublet,

$$\begin{aligned}\chi_L &= \nu'_L + \epsilon\,\tilde{\nu}_L \ , \\ \chi_R &= \tilde{\nu}_R + \epsilon\,\nu'_R \ ,\end{aligned} \tag{3.28}$$

with $\epsilon \sim y_\nu v/(Y_\nu v_\ell) \ll 1$.

In standard ADM models, the baryon and DM asymmetries are of similar size, depending on the exact form of the operators mediating them. Assuming the DM is relativistic at the decoupling temperature, this implies a DM candidate with a mass at $\sim \text{GeV}$ scale[1]. In particular, the relation between the DM mass and the relic abundances is given by

$$m_\chi = m_p \frac{\Omega_{\text{DM}}}{\Omega_{\text{B}}} \left| \frac{\Delta B}{\Delta \chi} \right| . \tag{3.29}$$

From Eqs. (3.25) and (3.26), $|\Delta B/\Delta \chi| \approx 1$ is obtained. This in turn yields a dark matter mass of

$$m_\chi \simeq 5 \text{ GeV} . \tag{3.30}$$

A mass of few GeV makes it difficult for the symmetric DM component to efficiently annihilate away. This is a generic challenge in ADM scenarios. This issue is circumvented by arranging for a light Higgs boson with $\sim$ GeV mass into which the DM can annihilate efficiently. Provided the light scalar has a significant CP-odd component, the Yukawa interactions in Eq. (3.5) and the DM content (3.28) imply that, to leading order in $\epsilon$, the coupling takes the form,

$$\mathscr{L}_{\text{DM}} \approx Y_\chi \bar{\chi} \gamma^5 \phi \chi . \tag{3.31}$$

This provides a natural DM annihilation channel as shown in Fig. 3.4. Using this interaction gives

$$(\sigma v)_{\text{NR}} = \frac{Y_\chi^4 m_\chi^6 v^2}{6\pi(2m_\chi^2 - m_\phi^2)^4} \left(1 - \frac{m_\phi^2}{m_\chi^2}\right)^{5/2} . \tag{3.32}$$

---

[1] Note that it is also possible to realize an ADM scenario with a DM mass of several TeV via Xogenesis [76]. In that case the DM mass is approximately ten times the decoupling temperature and must be roughly $m_{Z'}/2$ to have a sufficiently large annihilation cross section.

Writing the thermally averaged annihilation cross section as $\langle \sigma_A v \rangle = \sigma_0 \left(T/m_\chi\right)$ where

$$\sigma_0 = \frac{Y_\chi^4 m_\chi^6}{2\pi(2m_\chi^2 - m_\phi^2)^4} \left(1 - \frac{m_\phi^2}{m_\chi^2}\right)^{5/2}, \qquad (3.33)$$

the present energy density of the symmetric component of the DM particles is [77, 78]

$$\Omega_\chi h^2 \simeq \left(\frac{1.75 \times 10^{-10}}{\text{GeV}^2}\right) \frac{1}{\sigma_0 \sqrt{g_*}} \left(\frac{m_\chi}{T_f}\right)^2. \qquad (3.34)$$

Here $T_f$ is the freeze-out temperature and $g_*$ is the number of relativistic degrees of freedom as stated earlier.

For $m_\phi \approx 1\,\text{GeV}$, the remnant symmetric component will be subdominant to the asymmetric component produced by the $SU(2)_\ell$ phase transition provided the Yukawa coupling satisfies,

$$Y_\nu \gtrsim 0.1 \,. \qquad (3.35)$$

Light scalar bosons of mass $\sim$ GeV are somewhat unexpected from a potential whose overall energy scale is characterized by $v_\ell \sim$ TeV. However, they can be realized provided the quartic couplings are all small, and small quartic couplings are also favored by the need for a strongly first order phase transition. Additionally, large values of $\tan \beta$ indicate that $v_1 \gg v_2$ can provide such light masses. A detailed analysis of the scalar sector is beyond the scope of this chapter and this thesis, however it is a generic prediction of ADM models that there will be light ($\sim$ GeV) scalar particles with weak ($\sim 10^{-3}$) couplings to leptons [79, 80]. Such particles are typically not currently constrained by low energy experiments, but may be accessible in the future [81, 82].

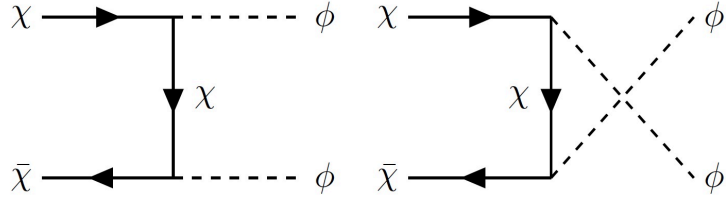

Figure 3.4: Dark matter annihilation channels to the pseudoscalar component of $\Phi_2$.

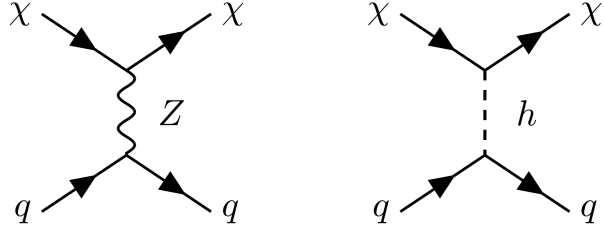

Figure 3.5: Diagrams contributing to DM interacting with quarks.

The coupling of the new scalar $\phi$ to the SM Higgs via the quartic $\lambda |\Phi|^2 |H|^2$ terms in the scalar potential allows for thermal equilibrium between $\phi$ and the SM provided the decay rate $\Gamma(\phi \to \mu\bar{\mu})$ is greater than the expansion rate at temperatures $T \sim 1$ GeV. This holds true for quartic coupling $\lambda \gtrsim 10^{-6}$. While the level of scalar mixing corresponding to the lower limit is too small to be observable at the LHC, larger values could be detectable.

The primary mediator for dark matter scattering with heavy nuclei is the $Z'$ gauge boson, which does not interact with quarks and therefore does not appear in tree-level DM direct detection diagrams although contributions do appear at one loop. As a result, the corresponding bound on $v_\ell$ set by the null search results from the CDMSlite experiment [83] is much less stringent than the collider constraint from CERN's LEP-II experiment of $v_\ell \gtrsim 1.7$ TeV. The DM direct detection diagrams involving the SM gauge bosons are shown in Fig. 3.5. The calculation of the spin-independent direct detection cross section closely follows the results of [64] for the electroweak diagrams, which, when combined with the CDMSlite bounds, provides an upper limit on the doublet admixture parameter $\epsilon \lesssim 0.3$, consistent with the assumption that $y'_{\nu,e} v \ll Y_{\ell,\nu,e} v_\ell$.

## 3.6 Conclusions

A new local symmetry beyond the Standard Model has been investigated in this chapter, wherein $SU(3)_C \times SU(2)_W \times U(1)_{\text{EM}}$ is supplemented by a non-Abelian gauge group $SU(2)_\ell$. The subscript $\ell$ corresponds to a generalization of lepton number as lepton number itself is anomalous. Standard Model leptons are promoted to doublets by introducing lepton partner fermions, all of which contain a new dark charge $\ell$. The doublet partners pair up with extra $SU(2)_\ell$ singlet fields needed to cancel the anomalies and develop vector-like masses after $SU(2)_\ell$ breaking. The global symmetries produce a viable dark matter candidate, which is the lightest of the lepton partners and stable due to a residual $U(1)_\chi$ global symmetry. It provides a viable baryogenesis mechanism through the breakdown of the new gauge group and naturally fits into an asymmetric dark matter framework.

The dynamics of an $SU(2)_\ell$ phase transition in the very early Universe are explicitly analyzed and a baryogenesis mechanism is found via nonperturbative interactions mediated by $SU(2)_\ell$ instantons. The instanton induced interactions violate both lepton and dark matter numbers and lead to a correlated asymmetry in both sectors.

There is a wide swath of parameter space which is relatively unconnected to electroweak observables to easily allow for a first order phase transition in the early Universe. Sufficient asymmetry is generated in the leptonic sector to give a baryon asymmetry via electroweak sphalerons which matches cosmological observations of the matter/antimatter asymmetry in Nature. Provided the mass of the dark matter is a few GeV, its correlated abundance will also match cosmological observations. This is critical as all observations up until today have shown a maximal matter/antimatter asymmetry [8] and a primordial asymmetry at the time of Big Bang Nucleosynthesis (temperatures of approximately 100 keV - 1 MeV) of $30,000,000$ matter particles to $30,000,001$ antimatter particles [78] in the quark sector.

Neither the lepton partners nor the additional neutral gauge bosons feel the strong force. They are colorless and thus are not efficiently produced at the LHC. The most stringent constraints are bounds on the $Z'$, which contributes to $e^+e^-$ at LEP-II. It would be interesting to see how a future high energy $e^+e^-$ collider could shed light on such a scenario, and whether it could preclude enough of parameter space to say something definitive about its potential realization as the mechanism for baryogenesis.

It may also be of interest to explore similar extensions of the Standard Model based on generalizations of baryon number. This is currently work in progress [84]. The most promising extension appears to be the promotion of the non-anamolous quantum number $B - L$ [85], another work in progress which is discussed further in Chapter 4 of this thesis. This is perhaps unsurprising as symmetries under both global and local transformations have thus far proven the most powerful in describing Nature.

This chapter and the previous chapter describe past research endeavors in great detail. I now move into the present and look towards the future, in somewhat lesser detail.

# Chapter 4

# Future Research

The published work of the previous chapters explored the ramifications of symmetry breakings and restorations in theoretical particle physics. This final chapter sketches progress on as yet unpublished work that has built upon those earlier projects. In some cases the relationship is obvious, in others it peripherally related, at times by a seemingly discontinous gap.

## 4.1 Old symmetries: Millicharged Scalars as Dark Matter

An in-depth look at the viability of the minicharged scalar proposed in Chapter 2 as a dark matter candidate. This works aims at near future and futuristic dark matter direct detection experiments [41].

### 4.1.1 Model outline

The current upper limits on the mass of the photon [32] are used in order to investigate the possibility that the electromagnetic interaction underwent a phase transition in the high temperature early Universe. The project of Chapter 2 investigated a similar scenario for a different cosmological epoch, namely a broken $U(1)_{\text{EM}}$ in the cosmological future, at temperatures colder than the cosmic background radiation today, $T \lesssim 2.7~K \approx 10^{-4}$eV. This model looks to the past and calculates the allowed parameter space for symmetry breaking in the higher temperature regime.

Constraints on the mass and charge of the scalar field used to break $U(1)_{\text{EM}}$ via the Higgs mechanism will be given and prospects for detection by new and proposed dark matter experiments such as LDMX [42], MilliQan [86], SHiP [23], FerMINI [87] and NA64 [22] will be investigated.

## 4.2 New symmetries: The BIND Model

Tightly linked to the research of Chapter 3, this project dispenses with anomalous symmetries altogether and is titled "BIND: Baryogenesis, Inflation, Naturalness/Neutrino Mass, and Dark Matter from a Non-Abelian Gauged $B - L$" [85].

### 4.2.1 Model outline

An extension of the Standard Model is constructed by promoting the baryon and lepton number difference, B-L, to a non-Abelian gauge symmetry. While both baryon number and lepton number are anomalous global symmetries, and in fact L may be broken at the electroweak phase transition via $SU(2)_W$ instantons, B-L is not anomalous and may be a good symmetry of Nature at all energies. The proposed $SU(2)_{B-L}$ is broken in two parallel models. The first breaks down to the global $U(1)_{B-L}$ via a Higgs field in the adjoint representation of $SU(2)$, the second breaks to no local symmetries via a Higgs field in the fundamental representation. Consequences of the SM extension are candidates for dark matter and the inflaton as well as a mechanism for baryogenesis and neutrino masses.

| Field | B-L | Weak | $U(1)_Y$ |
|---|---|---|---|
| $\hat{q}_L = \begin{pmatrix} q_L \\ \bar{q}_L \end{pmatrix}$ | 1/3 <br> -1/3 | ±1/2 <br> ±1/2 | 1/6 <br> −1/6 |
| $\hat{q}_R = \begin{pmatrix} q_R \\ \bar{q}_R \end{pmatrix}$ | 1/3 <br> -1/3 | 0 | 1/6 <br> −1/6 |
| $\hat{L}_L = \begin{pmatrix} L_L \\ \bar{L}_L \end{pmatrix}$ | -1 <br> 1 | ±1/2 <br> ±1/2 | −1/2 <br> 1/2 |
| $\hat{L}_R = \begin{pmatrix} L_R \\ \bar{L}_R \end{pmatrix}$ | -1 <br> 1 | 0 | −1/2 <br> 1/2 |
| $\hat{e}_R = \begin{pmatrix} e_R \\ \bar{e}_R \end{pmatrix}$ | -1 <br> 1 | 0 | −1/2 <br> 1/2 |
| $\hat{\nu}_R = \begin{pmatrix} \nu_R \\ \bar{\nu}_R \end{pmatrix}$ | -1 <br> 1 | 0 | −1/2 <br> 1/2 |
| $\hat{\Phi}_{B-L} = \begin{pmatrix} \phi^+ \\ \phi^0 \end{pmatrix}$ | 1 <br> -1 | 0 | 0 |

Figure 4.1: Minimal particle content and charges under the gauge symmetries $SU(2)_W \times U(1)_Y \times SU(2)_{B-L}$.

### 4.2.2 Novel particle content

The main physical difference in this model from the $SU(2)_\ell$ model of Chapter 3 is that the new interaction is felt by baryons as well as leptons since both are charged under the quantity $B - L$. The conceptual difference is its dependence upon a symmetry that holds already in the Standard Model both globally and locally.

Additionally, there is a possibility of a much smaller new particle content with this model and minimally adding to the known particles of Nature may be a desirable feature. My motivation in minimizing new degrees of freedom is the lack of both firm ($\sim 5\sigma$) and confirmed experimental evidence for any new particles as of September 2019.

Figure 4.1 displays one of three sample particle contents for the BIND model. The fields and charges for particles charged under $SU(2)_{B-L}$ use particle and antiparticle pairs of the same chirality as separate components of $SU(2)$ doublets.

The first type of particle content is minimal, in that the fewest possible new particles are proposed, only $\nu_R$ in addition to the SM particles and the symmetry breaking field. Figure 4.1 summarizes the minimal content along with the charges under the relevant symmetries.

Specifically, the notation is as follows for the SM fields,

$$q_L \equiv \begin{pmatrix} u_L \\ d_L \end{pmatrix}, \quad q_R \equiv \begin{pmatrix} u_R \\ d_R \end{pmatrix}, \quad L_{L/R} \equiv \begin{pmatrix} \nu_{L/R} \\ e_{L/R} \end{pmatrix}, \tag{4.1}$$

with generation index left implicit because the model treats particle families identically.

New degrees of freedom include only a scalar field to break the symmetry and a right handed neutrino with its doublet partner. This work in progress, motivated by my thesis research, appears at this point to yield all of the the physical phenomena in its title. We shall see how it ends up.

## 4.3 Old symmetries, new particles: Two scalar fields for Cosmic Dawn

Particle physics and cosmology combine in this work in which no new interactions are proposed but new particles are required. This work in progress is titled "Dark Matter Evolution Effects on the Hubble Rate at Cosmic Dawn."

### 4.3.1 Model motivation

The radio astronomy measurement of the 21 centimeter absorption profile at redshifts around $z \approx 18$ by the EDGES collaboration [88] is inconsistent with standard $\Lambda$CDM cosmology. Just as the Standard Model of particle physics contains all known interactions with the important exception of neutrino masses, so there is a Standard Model of cosmology [89]. Standard cosmology is based upon Einstein's general relativity and is specified by a small positive cosmological constant, $\Lambda$, and the existence of cold dark matter, CDM.

This project offers an explanation for the unexpected strength of the 21 cm line from the time in the early Universe when the stars first began to shine, known as Cosmic Dawn. While $T_{21} \approx -200$ mK is the $\Lambda$CDM predicted strength in milli-Kelvin, EDGES observed a depth of $T_{21} \approx -500$ mK. The formula for the brightness temperature of this line is given in terms of the CMB temperature, the temperature of the gas clouds (the whole Universe was a mainly hydrogen and helium gas cloud at this time), redshift and the optical depth of the gas cloud [90],

$$T_{21}(\text{K}) \approx \frac{\text{T}_{\text{gas}} - \text{T}_{\text{CMB}}}{1 + \text{z}} \, \tau. \tag{4.2}$$

In order to increase the magnitude of $T_{21}(\text{K})$ it is clear that either decreasing the temperature of the gas at the beginning of Cosmic Dawn or increasing the temperature of the CMB accomplishs the goal. All of the early literature focused on these options, mostly on the cooling of the gas and for good reason. An easy way to cool down a gas is by exposing it to a cold bath. Cold dark matter was apparently right there, more than 5 times as much mass in the CDM but thought not to have significant baryonic interactions. Modeling viable baryon-dark matter interactions to explain the signal continues to be an active area of research.

There is also a third possibility and that is to alter the optical depth of the gas cloud. In fact, the anomalous result may be explained with a modified Hubble rate in the optical depth of the hydrogen and helium gas clouds at these redshifts. The theoretical value of the brightness temperature of the 21 cm line, $T_{21}(K)$, depends inversely on the Hubble parameter at the redshifts under consideration, $15 \lesssim z \lesssim 20$, via the formula for the optical depth of the IGM early in the reionization era. The expansion rate H(z) for the matter dominated phase of the Universe, $z > 6$, is approximated by $H(z) \approx H_0 \Omega_m^{\frac{1}{2}}(1+z)^{\frac{3}{2}}$. The 2018 value of $\Omega_m$ as given by the Planck collaboration is $0.3156$ and is used in theoretical calculations of $T_{21}(K)$. If the optical depth of the IGM depended only on the baryonic matter density, $\Omega_b$, then using the Planck 2018 value of $\Omega_b h^2 = 0.0225$ in $T_{21}(K)$ yields an absorption depth of $2.5$ times the previous limit. This is the amplitude of the signal measured by EDGES. Models of a dark sector that can accommodate the signal are to be proposed.

### 4.3.2 Model sketch

The 21 cm line from neutral hydrogen is one of the most important diagnostic tools for astronomers, cosmologists and recently particle physicists interested in dark matter. One of the first indications of the spiral structure of our Milky Way galaxy was the 21 cm emission from the gas clouds in the arms ([91]). The EDGES result has intensified interest in hydrogen lines from gas clouds at Cosmic Dawn.

As the first stars formed out of hydrogen and helium gas, they began emitting ultraviolet (UV) radiation which excited the surrounding hydrogen atoms. When electrons in hydrogen atoms transition from the $n = 2$ to $n = 1$ orbital energy levels, 121.6 nm wavelength (the UV energy band) photons are emitted. These Lyman-$\alpha$ photons have the ability to trigger the nearly forbidden spin-flip transition ($\tau_{\text{sf}} = 1.1 \times 10^7$ years) of the proton and electron's aligned spins [92]. This "hyperfine" transition emits a 21 cm line in the rest frame of the hydrogen atom. This can and does occur in the very early stages of the Epoch of Reionization, when most of the hydrogen is not yet ionized by stellar radiation.

The Experiment to Detect the Global Epoch-of-Reionization Signature (EDGES) was designed to measure the 21 cm line excited by the first stars at Cosmic Dawn ([88]). The theoretical value of the absorption strength of the cosmic background radiation (CBR or CMB due to its peak in microwave wavelengths today) by the spin-flip excited hydrogen gas was worked out early on by George Field and others beginning in the late 1950s ([93]) and culminating in a study in 2004 [90]. The absorption feature in the data, measured as a brightness temperature with respect to the CMB temperature and centered on redshift $z \sim 18$, was more than twice as deep as the theory could account for, $T_{21}^{\text{data}} \approx \frac{5}{2} T_{21}^{\text{theory}}$.

Various explanations for the signal have been explored in the literature including interactions between the hydrogen gas of the IGM and dark matter, dark radiation, early dark energy effects, fuzzy dark matter and MOND. The explanation of the EDGES result proposed in this project modifies the Hubble rate in the equation for the optical depth of the gas clouds at these redshifts by changing the equation of state for dark matter.

The modification of $H(z)$ requires dark matter to evolve in three distinct phases. It begins as the cold dark matter of $\Lambda$CDM, evolving with non-relativistic $(1+z)^3$, then transitions to a radiation like evolution at some point prior to $z \sim 20$. During the EDGES window, $15 < z < 20$, only baryonic matter contributes to the Hubble rate. At some later time, dark matter cools down and transitions from a $(1+z)^4$ type evolution to $(1+z)^3$ again.

There are multiple ways to manage such an evolution history for dark matter, and one is offered here. The first phase transition is straightforward to model, but the second transition implies an equation of state that is not a straightforward transition. I call it the "re-cooling" problem.

The problem of re-cooling is analogous to the problem of re-heating in the post inflationary period in early Universe cosmology, in which the inflaton field $\phi$ has finished the acceleration phase of its evolution and upon terminating, leaves effectively all of its energy in the potential $V(\phi)$. The Universe at this point is in a super cooled state which must be "re-heated." The re-heating mechanism - by which the energy is transferred from $V(\phi)$ to the SM fields - is an active area of research, with a review given by [94].

The project outlined here will likely not see publication for two reasons, the first being that the surprising coincidence of the value of $\Omega_b$ reproducing the exact signal strength was noticed and published in 2018, although it was used not as a dark matter model constraint but as evidence for the non-existence of particle dark matter [95]. The second reason is that the exotic and time dependent equation of state was highly non-trivial for me to model. As a particle physicist, I wanted to work with a Lagrangian because I can predict things like cross-sections and lifetimes, decay rates. Two interacting scalar fields as dark matter candidates was my attempted path but the path is currently blocked. A very interesting paper with an oscillating equation of state for dark matter was posted in 2019 [96] which accomplishes the exotic equation of state with a single scalar field.

## 4.4 Old symmetries, old particles: Baryonic Dark Matter, or, The Transport of the Aim

A quantum chromodynamics (QCD) project motivated by the desire to solve the dark matter problem with no new interactions or particles, using only the strong nuclear force, is described in this section. Due to the constraints on dark matter emission of known force mediators, any baryonic matter put forth as a dark matter candidate must be in a special configuration of interaction phase space. One such candidate, first investigated as the H-dibaryon [97], consists of a strongly bound $uuddss$ quark system [98, 99]. This configuration of six quarks is special because it is a color singlet, an isospin scalar and electrically neutral, that is $|(ud)(us)(ds)\rangle$ with $n = 0, L = 0, s = 0, B = 2$, where $n$ is the radial quantum number, $L$ is orbital angular momentum, $s$ is diquark spin and $B$ is baryon number.

Thus the configuration ensures that its interactions via the three known forces of particle physics are suppressed and explains the lack of detection by non-gravitational experiments. In late 2018, I brought the hexaquark dark matter work to QCD expert Stan Brodsky at SLAC National Accelerator Laboratory in the hopes of predicting a mass and wavefunction diameter for the hexaquark, within the viable parameter space, using his methodology and results. He and his collaborators, in particular Guy de Téramond, had developed a set of techniques to probe QCD phenomena using Light Front Holographic QCD (LFHQCD) [100]. The predicted mass was found to be

$$M_{\text{hex}} = 1.65 \pm 0.06 \text{ GeV} \qquad (4.3)$$

with uncertainties given by the analysis of the orbital and radial excitations of the light hadron spectrum as well as the strange quark mass correction [101, 102, 103].

There are known issues with the hexaquark model including its possible decay channel to the deuteron, the nuclei of heavy hydrogen consisting of one proton and one neutron with mass $M_{\text{D}} = 1875.612928(12)$ MeV. Suppression of the decay rate may be possible by studying the compactness of the hexaquark and ensuring its wavefunction overlap with that of the deuteron is minimized [104] as well as estimating its mass to be slightly above the deuteron mass. Lack of dark matter direct detection results may be explained by dark matter co-rotating with Earth rather than behaving as a dark matter wind [104]. However, during the time period in which our collaboration was attempting to both understand and resolve any outstanding issues, we understood that a hexaquark composed only of up and down quarks is also allowed, the $ududud$ configuration, with a mass slightly less than the deuteron mass. The mass could not be shifted upward and the deuteron might decay to it.

This posed a problem we could not overcome because of the tightly constrained deuteron abundance measurements from Big Bang Nucleosynthesis [8]. Stan saw that the hexaquark could exist as an excited state of the deuteron and thus would be contained within its Fock space. It would not be a fundamental particle on its own (optimistically termed the hexon) but would be a virtual fluctuation hexaquark. This state would be contained within nuclear wavefunctions and could be analyzed with QCD alone. Guy de Tèramond and I calculated the $512$ term wavefunction for a hexaquark with spin zero, strong isospin zero and spatial angular momentum zero and found it to be symmetric, a situation forbidden by the spin-statistics theorem. It might be possible to break the theorem but that is a forbidding path and I don't think anyone other than myself considered it (they knew better). However, we could allow the hexaquark a color index, removing one of the antisymmetrizations from the wavefunction and giving it the correct statistics, but this leads to questions on its compactness and mass that are not easy to answer.

In a discussion with new collaborator Iván Schmidt and Stan, I suggested that the hexaquark behaves as a Cooper pair or set of Cooper pairs in a lattice structure simulating the nuclear environment. This part of the model is work in progress and may be discarded in the final version. The model appears viable and a paper is on its way to journal submission [105]. It provides a possible theoretical foundation for long standing nuclear physics mysteries including the European Muon Collaboration (EMC) effect [106, 107]. The initial two person collaboration has grown to five people and at the time of writing it appears that hexaquarks may form from a type of QCD diquark condensate along the lines of color superconductivity [108] or even super-insulators.

Originally an attempt to offer a candidate for particle dark matter without expanding the interactions or particles of the Standard Model, by using LFHQCD techniques, it is ending up as a possible solution to a long standing nuclear physics result using ordinary QCD in novel ways.

Perhaps the best part of the story is that experimental tests of the model may be carried out in the not too distant future. Diffractive dissociation of alpha particles upon nuclear targets would be a robust test [109]. The model would definitively be ruled out if our proposed jet observables are not seen.

## 4.5 Conclusion

The projects outlined in the final chapter of the thesis all flow from the ideas in the first chapter. The published work of the subsequent chapters as well as new results from particle physics and cosmology over the past two years have had a great influence on these works in progress. Looking upon this work as a whole, it occurs to me that one of the underlying themes - that of minimalism - may not be the way to proceed. If, to name one of the outstanding problems in physics that this work attempts to offer a solution to, dark matter is a set of particles with its own rich structure just as the known particles of the Standard Model have, then the minimalist approach is incorrect.

There may be a set of interactions, a set of local symmetries, that dark matter participates in, as well as a generational structure as mysterious as the visible one is to us now. In fact it may even be a more natural state of affairs, the existence of a complex dark sector. On the other hand, it may be that we simply do not understand the geometry and dimensions of spacetime yet. One of the most compelling features of Einstein's theory of general relativity is the inevitability of all gravitational interactions from his observation that gravity is geometry. With his theory we could understand one of the four forces of Nature simply by understanding how objects move on a variety of curved surfaces. Perhaps there are more curved surfaces than we know of, for objects to move within and upon, perhaps implying extra dimensions for such geometries and topologies to exist within.

To be explicit, there appear to be three options for the identity of dark matter, which has been implicitly defined in the thesis as the measured $\sim 26\%$ of the total energy density of the Universe. One, that new particles exist that either do not interact via known forces or interact weakly through known forces. Next, that the laws of gravity must be adjusted to accommodate the wealth of observational data for dark matter. Finally, by considering Einstein's result that gravity is geometry, the equations of general relativity may be left unchanged and geometry may be modified. Perhaps with extra dimensions, or in ways that we have not yet imagined.

The way forward is through experiment, as it ever was. New dark matter experiments are being proposed and funded as this thesis is being written. Astrophysical and cosmological observations are now more relevant than ever to the field of particle physics and we have dedicated facilities coming online to discover the identity of dark matter and dark energy, in particular the Large Synoptic Space Telescope or LSST [110]. Perhaps the most exciting of all is the push to discover the origin of neutrino masses [111], the only experimentally verified Beyond the Standard Model physics we have. This thesis offers two possible models of neutrino masses, both from new non-Abelian gauge theory extensions of the Standard Model. Regardless of whether Nature has chosen either of these particular paths (and I would bet that She has not) it is an extremely exciting time to enter the field.

*nity Summer Study on the Future of U.S. Particle Physics: Snowmass on the Mississippi (CSS2013): Minneapolis, MN, USA, July 29-August 6, 2013*, 2013.

# Appendices

## A  Boltzmann Diffusion in the early Universe

Particle diffusion occurs during the early Universe phase transition between the initial state false vacuum and the final state true vacuum just as in electroweak symmetry breaking of the Higgs vacuum. Quantum nucleated bubbles of true vacuum energy expand outward into the symmetric region of zero vacuum energy, with particles in the bubble wall diffusing out into this symmetric state.

The rates at which the particle number densities change as they diffuse out are described by the following 12 equations. The common interaction rates for the $SU(2)_\ell$ instanton induced interactions for particle species $i$ are denoted by $\Gamma_i$. Time derivatives are labeled with a dot, spatial derivatives are taken with respect to coordinate $z$ normal to the surface of the bubble wall and labeled with a prime. The curvature of the bubble wall is assumed to be negligible (the wall is approximated to be flat) as we effectively zoom into a small expansion region [72].

Neglecting the neutrino and neutrino partner Yukawas, the equations for the particle number densities $n(l), n(\tilde{l})$ and $n(l')$ are given in simplified notation (all particle number densities labeled by their species, e.g. $\tilde{\ell} \equiv n(\tilde{\ell})$) by

$$\dot{l} - D_l \nabla^2 l = -\Gamma \left[ 4l + 2\tilde{l} - 4(\nu + e) - 2(\tilde{\nu} + \tilde{e}) \right] - \Gamma_{Y_l} \left( \frac{l}{2} - \frac{\Phi^u}{2} - \frac{l'}{2} \right) -$$
$$\Gamma_{y_e} \left( \frac{l}{2} - \frac{h}{4} - e \right), \tag{A.1}$$

$$\dot{\tilde{l}} - D_{\tilde{l}} \nabla^2 \tilde{l} = -\Gamma \left[ 4\tilde{l} + 2l - 4(\tilde{\nu} + \tilde{e}) - 2(\nu + e) \right] - \Gamma_{Y_l} \left( \frac{\tilde{l}}{2} - \frac{\Phi^d}{2} - \frac{l'}{2} \right) -$$
$$\Gamma_{y_e} \left( \frac{\tilde{l}}{2} - \frac{h}{4} - \tilde{e} \right), \tag{A.2}$$

$$\dot{l}' - D_{l'} \nabla^2 l' = -\Gamma_{y'_e} \left( \frac{l'}{2} - \frac{h}{4} - e' \right) + \Gamma_{Y_l} \left( \frac{l}{2} - \frac{\Phi^u}{2} - \frac{l'}{2} \right) + \Gamma_{Y_l} \left( \frac{\tilde{l}}{2} - \frac{\Phi^d}{2} - \frac{l'}{2} \right). \tag{A.3}$$

For $\nu, \tilde{\nu}, \nu'$,

$$\dot{\nu} - D_\nu \nabla^2 \nu = -\Gamma_{Y_\nu} \left( \nu - \frac{\Phi^u}{2} - \nu' \right) - \Gamma \left( 3\tilde{\nu} + \tilde{e} + 2e - l - 2\tilde{l} \right), \tag{A.4}$$

$$\dot{\tilde{\nu}} - D_{\tilde{\nu}} \nabla^2 \tilde{\nu} = -\Gamma_{Y_\nu} \left( \tilde{\nu} - \frac{\Phi^d}{2} - \nu' \right) - \Gamma \left( 3\nu + e + 2\tilde{e} - \tilde{l} - 2l \right), \tag{A.5}$$

$$\dot{\nu}' - D_{\nu'} \nabla^2 \nu' = \Gamma_{Y_\nu} \left( \nu - \frac{\Phi^u}{2} - \nu' \right) + \Gamma_{Y_\nu} \left( \tilde{\nu} - \frac{\Phi^d}{2} - \nu' \right). \tag{A.6}$$

For $e, \tilde{e}, e'$,

$$\dot{e} - D_e\nabla^2 e = -\Gamma_{Y_e}\left(e - \frac{\Phi^u}{2} - e'\right) + \Gamma_{y_e}\left(\frac{l}{2} - \frac{h}{4} - e\right) -$$
$$\Gamma\left(3\tilde{e} + \tilde{\nu} + 2\nu - l - 2\tilde{l}\right), \tag{A.7}$$

$$\dot{\tilde{e}} - D_{\tilde{e}}\nabla^2 \tilde{e} = -\Gamma_{Y_e}\left(\tilde{e} - \frac{\Phi^d}{2} - e'\right) + \Gamma_{y_e}\left(\frac{\tilde{l}}{2} - \frac{h}{4} - \tilde{e}\right) -$$
$$\Gamma\left(3e + \nu + 2\tilde{\nu} - \tilde{l} - 2l\right), \tag{A.8}$$

$$\dot{e}' - D_{e'}\nabla^2 e' = \Gamma_{y'_e}\left(\frac{l'}{2} - \frac{h}{4} - e'\right) + \Gamma_{Y_e}\left(e - \frac{\Phi^u}{2} - e'\right) + \Gamma_{Y_e}\left(\tilde{e} - \frac{\Phi^d}{2} - e'\right). \tag{A.9}$$

Finally, for the particle number densities $\Phi^u, \Phi^d$ and $h$,

$$\dot{\Phi}^u - D_{\Phi^u}\nabla^2 \Phi^u = \gamma_1 + \Gamma_{Y_l}\left(\frac{l}{2} - \frac{\Phi^u}{2} - \frac{l'}{2}\right) + \Gamma_{Y_\nu}\left(\nu - \frac{\Phi^u}{2} - \nu'\right) +$$
$$\Gamma_{Y_e}\left(e - \frac{\Phi^u}{2} - e'\right), \tag{A.10}$$

$$\dot{\Phi}^d - D_{\Phi^d}\nabla^2 \Phi^d = \gamma_2 + \Gamma_{Y_l}\left(\frac{\tilde{l}}{2} - \frac{\Phi^d}{2} - \frac{l'}{2}\right) + \Gamma_{Y_\nu}\left(\tilde{\nu} - \frac{\Phi^d}{2} - \nu'\right) +$$
$$\Gamma_{Y_e}\left(\tilde{e} - \frac{\Phi^d}{2} - e'\right), \tag{A.11}$$

$$\dot{h} - D_h\nabla^2 h = \Gamma_{y_e}\left(\frac{l}{2} - \frac{h}{4} - e\right) + \Gamma_{y_e}\left(\frac{\tilde{l}}{2} - \frac{h}{4} - \tilde{e}\right) - \Gamma_{y'_e}\left(\frac{l'}{2} - \frac{h}{4} - e'\right). \tag{A.12}$$

where $\gamma_1$ and $\gamma_2$ are the CP-violating sources.

## A.1 Constraints on particle number densities

The equilibrium conditions emerging from the Yukawa terms are given by

$$\frac{l}{2} - \frac{\Phi^u}{2} - \frac{l'}{2} = 0 , \quad \frac{\tilde{l}}{2} - \frac{\Phi^d}{2} - \frac{l'}{2} = 0 , \quad \nu - \frac{\Phi^u}{2} - \nu' = 0 , \quad \tilde{\nu} - \frac{\Phi^d}{2} - \nu' = 0 ,$$

$$e - \frac{\Phi^u}{2} - e' = 0 , \quad \tilde{e} - \frac{\Phi^d}{2} - e' = 0 , \quad \frac{l}{2} - \frac{h}{4} - e = 0 , \quad \frac{\tilde{l}}{2} - \frac{h}{4} - \tilde{e} = 0,$$

$$\frac{l'}{2} - \frac{h}{4} - e' = 0 . \tag{A.13}$$

The equilibrium conditions emerging from the instanton-induced interactions are

$$l - \nu - e = 0 , \quad \frac{l}{2} + \frac{\tilde{l}}{2} - \nu - \tilde{e} = 0 , \quad \tilde{l} - \tilde{\nu} - \tilde{e} = 0 , \quad \frac{l}{2} + \frac{\tilde{l}}{2} - \tilde{\nu} - e = 0 . \tag{A.14}$$

As discussed in Chapter 3, two of the Yukawa equilibrium conditions and three of the instanton conditions are linearly dependent on the others. Therefore, there are only four independent particle number densities. For example one can choose them to be $l$, $\tilde{\nu}$, $h$ and $\Phi^u$. In this case the other particle densities are given by

$$\tilde{l} = 2\tilde{\nu} - \frac{h}{2} , \quad , l' = l - \Phi^u , \quad \nu = \frac{l}{2} + \frac{h}{4} , \quad \nu' = \frac{l}{2} - \frac{\Phi^u}{2} + \frac{h}{4} ,$$

$$e = \frac{l}{2} - \frac{h}{4} , \quad \tilde{e} = \tilde{\nu} - \frac{h}{2} , \quad e' = \frac{l}{2} - \frac{\Phi^u}{2} - \frac{h}{4} , \quad \Phi^d = 2\tilde{\nu} - l + \Phi^u - \frac{h}{2} . \tag{A.15}$$

"For every symmetry there comes a constraint.
If physics is to look the same when the origin of time is shifted
[O]nly those processes that conserve energy are allowed.
If physical law is to be immune to the arbitrary displacement of our spatial axes,
then nature requires the conservation of linear momentum.
If the laws are to be unaffected by the arbitrary rotation of a coordinate system,
then angular momentum must be conserved.
If the laws are to be the same for all inertial observers,
then the space-time interval must be invariant.
[A]nother constraint...so beautiful as to make one jaw drop in wonder
Symmetry creates force.
[T]he symmetry of identical particles forces matter...to be enrolled as either fermion or boson
Bosons, typified by the photon, carry the fundamental forces that cause fermions to attract and repel.
Fermions, led by electrons and quarks, become constituents of ordinary matter.
Gravity. Electromagnetism. The strong force. The weak force.
Each fundamental interaction is called into being
by the requirements of a particular local symmetry."

*Michael Munowitz, author, PhD in Chemical Physics.* Knowing: The Nature of Physical Law, 2005.

Figure A.1: Emmy Noether's theorem for the layperson